\documentclass[pre,twocolumn,amsmath,amssymb,nofootinbib,floatfix,superscriptaddress]{revtex4}

\usepackage[normalem]{ulem}
\usepackage{graphicx, bm, tikz}
\usepackage[breaklinks]{hyperref}

\makeatletter
\def\graphicscale{\twocolumn@sw{0.3}{0.4}}
\def\graphicthreescale{\twocolumn@sw{0.3}{0.4}}

\begin{document}

\title{Dynamic scaling behavior in the presence of a periodic magnetic driving\\
  across Ising continuous transitions}

\author{Andrea Pelissetto}
\affiliation{Dipartimento di Fisica dell'Universit\`a di Roma
  ``La Sapienza" and INFN, Sezione di Roma I, I-00185 Roma, Italy}

\author{Ettore Vicari} 
\affiliation{Dipartimento di Fisica dell'Universit\`a di Pisa,
  Largo Pontecorvo 3, I-56127 Pisa, Italy}

\date{\today}

\begin{abstract}
  We study the critical dynamics arising from a time-dependent
  periodic homogenous source coupled to the order-parameter field,
  which drives a classical ferromagnetic system across a continuous
  transition. For this purpose, we consider the paradigmatic
  two-dimensional (2D) Ising model in the presence of a periodic
  magnetic field $h(t)=-A\, \cos (2\pi t/P)$, evolving under a purely
  relaxational dynamics at the critical temperature.  We show that the
  periodic driving gives rise to a peculiar dynamic scaling behavior
  in the thermodynamic limit, arising from a nontrivial interplay
  among the time $t$, the amplitude $A$ and period $P$ of $h(t)$. The
  relevant scaling variables are $\tau=t/P$ and $\sigma=A P^\kappa$,
  with $\kappa = y_h/z$, where $y_h=(d+2-\eta)/2$ is the critical
  dimension of the magnetic field, and $z$ is dynamic exponent for the
  critical relaxational dynamics ($\kappa\approx 0.865$ for the 2D
  Ising model).  The dynamic scaling behaviors of the magnetization and
  bond-energy density show an oscillatory behavior around a smooth
  curve which approaches a large-$\tau$ stationary behavior.  We also
  briefly discuss the dynamic behavior of an Ising system driven
  across the critical point by a periodic time-varying temperature at
  zero magnetic field.
\end{abstract}

\maketitle

\section{Introduction}
\label{intro}

The long-range critical modes at continuous phase transitions exhibit
well-known equilibrium and out-of-equilibrium scaling behaviors; see,
e.g.,
Refs.~\cite{Wilson-83,WK-74,Fisher-74,Wegner-76,Aharony-76,Ma-book,
  Binder-76,HH-77,Kibble-80,Zurek-96,PV-02,Cardy-book,CG-05,FM-06,
  Sachdev-book,CEGS-12,PV-16,DKPR-16,Biroli-16,PRV-18,RV-21,TV-22,DV-23,BPV-25}
for a (non-exaustive) list of references.  In this paper we extend the
phenomenology of the universal dynamic scaling at thermal continuous
transitions, by studying how a periodic external driving affects the
critical dynamic behavior.  Specifically, we analyze---both
conceptually and numerically in Ising systems---the dynamic scaling
that develops when a periodic source couples to the order-parameter
field, considering a purely relaxational dynamics without conservation
laws~\cite{Ma-book} (model-A critical dynamics in the classification
of Ref.~\cite{HH-77}).

Periodic time-dependent couplings have already been considered at
zero-temperature quantum transitions~\cite{Sachdev-book,RV-21} in
quantum many-body systems, driven by the quantum unitary dynamics,
see, e.g.,
Refs.~\cite{GH-98,SAN-10,RSS-12,LMPPA-17,BDP-17,RDS-21,RRDOR-23,
  CUBZL-23,TS-23,CZQXX-25,SG-26,YTZL-26,GPJ-26}.  Instead, the role of
a periodic driving at thermal phase transitions (driven by thermal
fluctuations) has received much less attention. Some studies have
focused on searching for dynamic transitions unrelated from the
underlying equilibrium transition, induced by the periodic driving in
ferromagnetic systems, see, e.g.,
Refs.~\cite{LP-90,SRN-98,CA-99,PP-13,MBR-24,BMR-25,XKM-26}.  In this
work we focus on periodic time-dependent couplings that drive the
system across the continuous transitions and show that, within the
critical region, the system obeys general scaling laws, which can be
derived using standard renormalization-group (RG) arguments.  In
particular, we show that the singular scaling behavior is controlled
by the dynamic universality class of the transition in equilibrium, so
the dynamic scaling exponents can be directly related to the critical
exponents of the underlying continuous transition.

In this exploratory study of the effects of a periodic driving across
a classical continuous ferromagnetic transition, we consider the
paradigmatic two-dimensional (2D) Ising model, for which several exact
results are available, for instance, the critical temperature and the
universal critical exponents~\cite{Onsager-44,FF-69}.  We focus on the
dynamic behavior of the system, evolving under a standard Metropolis
dynamics ~\cite{Metropolis:1953am,Binder-76}---an example of purely
relaxational dynamics---in the presence of a periodic magnetic field
at the critical temperature. We identify the emerging dynamic scaling
behavior, characterized by an oscillatory synchronized time-dependent
magnetization, which approaches an asymptotic large-time stationary
behavior.  These scaling features are expected to be quite general,
and therefore they should also characterize Ising models in higher
dimensions, and also continuous transitions of generic ferromagnetic
systems driven by a periodic external field coupled to the order
parameter at the critical temperature.  We also briefly discuss the
scaling dynamic behavior when the temperature varies periodically at
zero magnetic field, driving again the system across the critical
transition.

The paper is organized as follows. In Sec.~\ref{twodisi} we present
the 2D Ising model and summarize the relevant RG properties at its
continuous transition: the static and dynamic exponents and the RG
dimensions of the relevant operators. In Sec.~\ref{isidriv} we focus
on the behavior in the presence of a periodic magnetic field at the
critical temperature: in Sec.~\ref{perh} we discuss the protocol,
while in Sec.~\ref{dynsca} we put forward a dynamic scaling theory for
the time dependence of observables such as the magnetization and the
bond-energy density.  In Sec.~\ref{numres} we present a numerical
analysis of the 2D Ising model subject to a periodic magnetic driving.
In Sec.~\ref{tempdriv} we discuss the dynamic scaling arising in the
presence of a periodically varying temperature that drives the system
across the transition point at zero magnetic field.  Finally, in
Sec.~\ref{conclu} we summarize and draw our conclusions.

\section{The model}
\label{twodisi}

We consider the square-lattice 2D Ising model with
Hamiltonian
\begin{eqnarray}
  H(J,h) = - J \sum_{\langle
    {\bm x} {\bm y} \rangle} s_{\bm x} s_{\bm y} - h
  \sum_{\bm x} s_{\bm x} ,
\label{classisi}  
\end{eqnarray}
where $s_{\bm x}=\pm 1$ are classical spin variables, $h$ is an
external homogenous magnetic field, the first sum is over all lattice
bonds ${\langle {\bm x} {\bm y} \rangle}$, and the second one is over
all lattice sites ${\bm x}$.  The partition function is given by
\begin{equation}
Z = \sum_{\{s_{\bm x}\}} \exp[-\beta H(J,h)],\qquad \beta=1/T.
\label{zfunc}
\end{equation}
In our numerical work
we consider square systems of size $L$ with
periodic boundary conditions and set $J=1$ without loss of generality.

The square-lattice Ising model (\ref{classisi}) undergoes a continuous
transition at $h=0$ and $T_c=2/\ln(\sqrt{2}+1)$~\cite{Onsager-44},
with exponents $\nu=1$ (controlling the divergence of the length scale
of the critical modes) and $\eta=1/4$ (controlling the power-law decay
of the critical spin-spin correlation function); see, e.g.,
Ref.~\cite{Ma-book}. They are related to the RG dimension $y_t$ of the
even RG perturbation associated with the (reduced) temperature by
$y_t=1/\nu=1$, and to the odd RG perturbation associated with the
field $h$ by $y_h = 2-\eta/2=15/8$; see, e.g., Ref.~\cite{PV-02}.  The
RG dimension $y_\phi$ of the order-parameter field is given by $y_\phi
= 2- y_h = \eta/2=1/8$, while the RG dimension $y_e$ of the energy
density is given by $y_e = 2- y_t = 1$. The critical dynamics is
characterized by the universal dynamic exponent $z$ (controlling the
autocorrelation time of the critical modes), which depends both on the
static universality class and on the nature of the
dynamics~\cite{Ma-book,HH-77}. We consider a purely relaxational
dynamics (model A dynamics in the classification of
Ref.~\cite{HH-77}), for which $z$ has been accurately estimated
numerically~\cite{NB-00,WH-97,NB-96,G-95,CV-11}, obtaining\footnote{We
report here some recent estimates of the dynamic exponent $z$, mainly
obtained in Monte Carlo studies under equilibrium conditions:
$z=2.1667(5)$ (Ref.~\cite{NB-00}), $z=2.168(5)$ (Ref.~\cite{WH-97}),
$z=2.1665(12)$ (Ref.~\cite{NB-96}), $z=2.172(6)$ (Ref.~\cite{G-95}),
$z=2.170(6)$ (Ref.~\cite{CV-11}).  The estimate $z=2.167(1)$ takes
into account the above results.}  $z=2.167(1)$.

\section{Periodic magnetic driving at the critical temperature}
\label{isidriv}

\subsection{Dynamic protocol}
\label{perh}

We consider a dynamic protocol in which the magnetic field
$h(t)$ varies periodically around $h=0$ and the temperature is fixed 
at $T_c$. More precisely, we consider 
\begin{equation}
  h(t) = - A \cos\left( {2\pi t\over P}\right), 
  \label{htdef}
\end{equation}
with amplitude $A>0$ and period $P$.  The dynamics starts at $t=0$
from thermalized configurations at $h_i = -A < 0$ and $T=T_c$, thereby
breaking the ${\mathbb Z}_2$ parity invariance. This allows us to also
investigate if, and how, the ${\mathbb Z}_2$ symmetry is recovered in
the large-time limit.

The time evolution is obtained by performing standard Metropolis local
updates~\cite{Metropolis:1953am,Binder-76}. At each site, spin flips
are performed with probability
\begin{equation}
P(s_{{\bm x}}\to -s_{{\bm x}}) = {\min}(1,e^{-\Delta H}),
\label{metroup}
\end{equation}
where $\Delta H$ is the change of the Hamiltonian when
replacing $s_{{\bm x}}$ with $ -s_{{\bm x}}$.  In our implementation,
spins are updated using a checkerboard scheme. Since the square lattice
is bipartite, sites can be divided in two sets, the set of even and
odd sites, respectively. We first update all spins at even sites, then
all spins at odd sites.  A time unit corresponds to a complete lattice
sweep.

The evolution of the system is monitored by the time-dependent
magnetization defined as
  \begin{eqnarray}
M(t) \equiv {1\over L^2} \sum_{\bm x} m_{\bm x}(t),
\quad
m_{\bm x}(t) \equiv \langle s_{\bm x} \rangle_t,
\label{magnt2d}
  \end{eqnarray}
  and the subtracted bond-energy density
  \begin{eqnarray}
    &&E_s(t) \equiv E(t) - E_c,
    \label{esdef}\\
  &&E(t) = 
  {1\over 2L^2}  \langle \sum_{\langle
    {\bm x} {\bm y} \rangle} s_{\bm x} s_{\bm y} \rangle_t,
  \qquad E_c = \sqrt{2},
\label{esdeft}
\end{eqnarray}
  where $E_c$ is the equilibrium expectation value of the bond-energy
  density at the critical point~\cite{Onsager-44,FF-69}.  Since the
  dynamic protocol involves a stochastic relaxational process, results
  are obtained by averaging over the trajectories starting from the
  Gibbs ensemble of thermalized configurations at $h_i$ and $T=T_c$.

\subsection{Dynamic scaling theory}
\label{dynsca}

In this section we put forward a dynamic scaling theory for the time
dependence of global observables in critical ferromagnetic Ising
systems under a periodic magnetic field.  For simplicity, we mainly
focus on the scaling behavior for $T=T_c$, as in the dynamic protocol
presented above, but we will also briefly mention how to generalize
the scaling predictions for $T\not= T_c$, but still in the critical
region.

We first discuss the scaling theory for finite-size systems,
generalizing the dynamic finite-size scaling (FSS)
theory~\cite{FB-72,Barber-83,Privman-90,PV-02,Cardy-book,CPV-14,
  PRV-18,RV-21,TV-22,DV-23} developed for systems with a
time-independent magnetic field.  A reasonable hypothesis is that the
amplitude $A$ of the magnetic field $h(t)$ scales as the magnetic
field in the time-independent case, while the period $P$ scales as the
time variable $t$.  Therefore, the relevant scaling variables for the
dynamics at the critical temperature $T_c$ are expected to be
\begin{equation}
  W_a = A L^{y_h},\qquad
  W_t = t L^{-z}, \qquad
  W_p = P L^{-z}, 
  \label{ufssdef}
\end{equation}
where $y_h$ is the RG dimension of the magnetic field
and $z$ is the dynamic critical exponent for the dynamics.
Dynamic FSS is obtained in the 
limit $A\to 0$, $t\to\infty$, $P\to\infty$ and $L\to\infty$
keeping the FSS variables $W_{\#}$ fixed.  In this limit the
time-dependent magnetization is expected to behave as
\begin{equation}
  M(t,A,P,L)\approx L^{-y_\phi} {\cal M}_L(W_a,W_t,W_p),
    \label{mtfss}
\end{equation}
where $y_\phi$ is the RG dimension of the order-parameter field, while 
the subtracted bond-energy density $E_s$ defined in Eq.~(\ref{esdef})
should behave as 
\begin{equation}
  E_s(t,A,P,L)\approx L^{-y_e} {\cal E}_L(W_a,W_t,W_p), 
    \label{erfss}
\end{equation}
where $y_e$ is the RG dimension of the energy density. 

Scaling relations in the thermodynamic limit can be straightforwardly
obtained by taking the $L\to\infty$ limit keeping the scaling
variables $\sigma$ and $\tau$, defined as
\begin{eqnarray}
  \sigma &=& W_a \, W_p^{y_h/z} = A P^{\kappa}, \qquad \kappa = y_h/z,
  \label{whdef}\\
  \tau &=& W_t \, W_p^{-1} = t/P, \label{wpdef}
\end{eqnarray}
fixed. For the 2D Ising transition, using
$y_h=15/8$ and $z=2.167(1)$, we obtain
\begin{equation}
      \kappa = 0.8653(4).
      \label{kappaval}
\end{equation}
Then, we obtain
\begin{eqnarray}
  M(t,A,P)&\approx& A^{\zeta} {\cal M}(\sigma,\tau),
  \qquad \zeta = y_\phi/y_h,
    \label{mtthlim} \\
  E_s(t,A,P) &\approx& A^{\varepsilon} {\cal E}_s(\sigma,\tau),
  \qquad \varepsilon = y_e/y_h.
    \label{enethlim}
\end{eqnarray}
At the 2D Ising transition $\zeta = 1/15$ and $\varepsilon =
8/15$. These scaling expressions hold in the thermodynamic limit, for
$A\to 0$, $t,P\to\infty$ keeping $\sigma$ and $\tau$ fixed.  The
dynamic scaling functions are expected to be universal, apart from a
multiplicative normalization and the normalizations of the scaling
variables.

If we consider protocols at fixed temperature $T\not= T_c$ in a finite
volume, we should define an additional scaling variable $W_\delta =
\delta T \,L^{y_t}$, where $y_t=1/\nu$ is the RG dimension associated
with $\delta T = T-T_c$.  In the infinite-volume limit, the relevant
variable is the ratio $W_\delta/W_p^{\lambda} = \delta T\,
P^{\lambda}$ with $\lambda=y_t/z$ [for the 2D Ising transition
  $\lambda=0.4615(2)$].  This implies that, to observe the dynamic
scaling behavior, $T$ should approach $T_c$ as $P \to \infty$, keeping
$\delta T P^{\lambda}$ constant.

The asymptotic dynamic scaling behavior is approached with power-law
suppressed corrections. Under equilibrium conditions, scaling
corrections decay as $\xi^{-\omega}$, where $\xi$ is the diverging
correlation length (at $T_c$, it diverges as $h^{-1/y_h}$ for $h\to
0$) and $\omega$ is the leading scaling correction
exponent~\cite{PV-02} ($\omega=2$ for the 2D Ising universality
class~\cite{PV-02,CHPV-02}). In the dynamic case, we expect
corrections decaying as $A^{\omega/y_h}=A^{16/15}$ or, equivalently,
as $P^{-\omega/z} = P^{-0.92}$.  Scaling corrections may also arise
from the breaking of time-translation invariance due to the ${\mathbb
  Z}_2$-breaking initial condition of the dynamic protocol.  It is
plausible that they decay as $t^{-1}\sim P^{-1}$ in the dynamic
scaling limit, in analogy with what happens in finite-size systems,
where corrections behave as $L^{-1}$ in the presence of boundaries
that break translation invariance ~\cite{CPV-14}.

The scaling functions can be determined in some particular limits.
For $t=\tau=0$, the system is in equilibrium at $h=-A$ and $T=T_c$.
Therefore, the magnetization takes its equilibrium value $M_{\rm
  eq}(h,T_c)$. For $h\to 0$, we have
\begin{equation}
M_{\rm eq}(h,T_c) \approx {\rm sgn}(h)\, c\,|h|^\zeta,\quad c = 1.00268751,
  \label{mequ}
\end{equation}
where the constant $c$ has been obtained using the results reported in
Refs.~\cite{CH-00,Wu-66}.  Therefore,
\begin{equation}
  {\cal M}(\sigma,\tau=0)=-c,
  \label{mtau0}
  \end{equation}
independently of $\sigma$.  

Scaling functions can be computed in the limit $\sigma\to\infty$.
Since the period $P$ is much larger than the typical autocorrelation
time of the slowest modes, which scales as $A^{z/y_h}$, the system is
always effectively in equilibrium.  Note that this statement is not in
contrast with the fact that any statistical system is out of
equilibrium when crossing a continuous transition. Indeed, one can
show that the out-of-equilibrium behavior occurs in a time interval
$\Delta t$ that scales as $P \sigma^{-\mu}$, $\mu = 1/(1+\kappa)$, so
the rescaled $\Delta \tau =\Delta t/P$ vanishes for $\sigma \to
\infty$.  To prove this result, let us note that, close to a crossing
time $t_{\rm cr}$ where $h(t_{\rm cr}) = 0$, for $P\to\infty$ we have
$h(t) = \tilde{t}/t_s$ with $t_s = P/A$, $\tilde{t} = t - t_{\rm cr}$.
Since $t_s\to \infty$ in the scaling limit, the behavior is analogous
to that of a statistical system subject to a Kibble-Zurek (KZ)
dynamics~\cite{Kibble-80,Zurek-96,CEGS-12,PV-16,Biroli-16,RV-21,TV-22}.
In the KZ case the relevant scaling variable is $h(t)
\tilde{t}^{y_h/z} = \tilde{t}^{1/\mu}/t_s$ or, equivalently, the ratio
$\tilde{t}/t_s^{\mu}$.  The out-of-equilibrium behavior occurs in a
finite interval of the ratio $\tilde{t}/t_s^{\mu}$, i.e., in a time
interval $\Delta t \sim t_s^\mu = P \sigma^{-\mu}$, as claimed above.

Since the system is always in equilibrium, for $\sigma\to \infty$ we 
expect $M(t)\approx M_{\rm eq}[h(t)]$. 
Therefore, using Eq.~(\ref{mequ}), we predict
\begin{eqnarray}
  {\cal M}(\sigma\to\infty,\tau) = {\rm
    sgn}[h(t)]\,c\,|\cos(2\pi\tau)|^\zeta.
  \label{sigmainfi2}
\end{eqnarray}
Finally, the scaling functions can be predicted for $\sigma\to 0$. In
this case $P$ is very small compared with the typical correlation time
of the dynamics that scales as $A^{z/y_h}$.  Therefore, the
magnetization does not have enough time to change, so we expect
$M(t,h,p) \approx M_{\rm eq}(h_i)$, thus
\begin{eqnarray}
  {\cal M}(\sigma\to 0,\tau) = - c.
  \label{sigma02}
\end{eqnarray}
One may also consider the dynamic protocol far from criticality, at
fixed temperature $T\neq T_c$.  For $T > T_c$ we expect to observe an
effective equilibrium behavior in the limit $t,\,P\to \infty$ keeping
$\tau=t/P$ fixed, for any amplitude $A$, so
\begin{equation}
  M(t,A,P,T))\approx M_{\rm eq}[h(t),T].
  \label{ntlTc}
\end{equation}
The behavior is more complex for $T<T_c$, since the periodic magnetic
field makes the system periodically cross the first-order transition
line. Some investigations have been already reported in the
literature; see, e.g.,
Refs.~\cite{LP-90,SRN-98,CA-99,PP-13,MBR-24,BMR-25}.

We finally mention that the above dynamic scaling arguments can also
be applied to quantum systems driven across a quantum transition by a
periodic magnetic field (in this case one would consider the quantum
unitary dynamics of closed systems, for which $z=1$).  Some results
for round-trip protocols in quantum Ising systems subject to a
time-dependent longitudinal field are reported in Ref.~\cite{TV-22}.

\section{Numerical results for a magnetic driving}
\label{numres}

\subsection{Average behaviors over trajectories}

We now present a numerical analysis of the protocol outlined in
Sec.~\ref{perh}, showing the emergence of the dynamic scaling behavior
outlined in Sec.~\ref{dynsca}. For this purpose, we determine the
average magnetization and bond energy as a function of time, averaged
over at least (approximately) 100 trajectories, for several values of
$P$ and $A$, and for sizes $L$ ranging from 300 to 800.  We analyze
the behavior of the rescaled magnetization and subtracted bond-energy
density, respectively ${\cal M}(\sigma,\tau)\approx A^{-\zeta} M(t,A,P)$
and ${\cal E}_s(\sigma,\tau) \approx A^{-\varepsilon} E_s(t,A,P)$, at
fixed $\sigma=A P^\kappa$.

%%%%%%%%%%%%%%%%%%%%%%%%%%%%%%%%%%%%%%%%%%%%%%%%%%%%%%%%%%%%%%%%%%%%%%%%
\begin{figure}[!t]
 \includegraphics*[scale=\graphicscale]{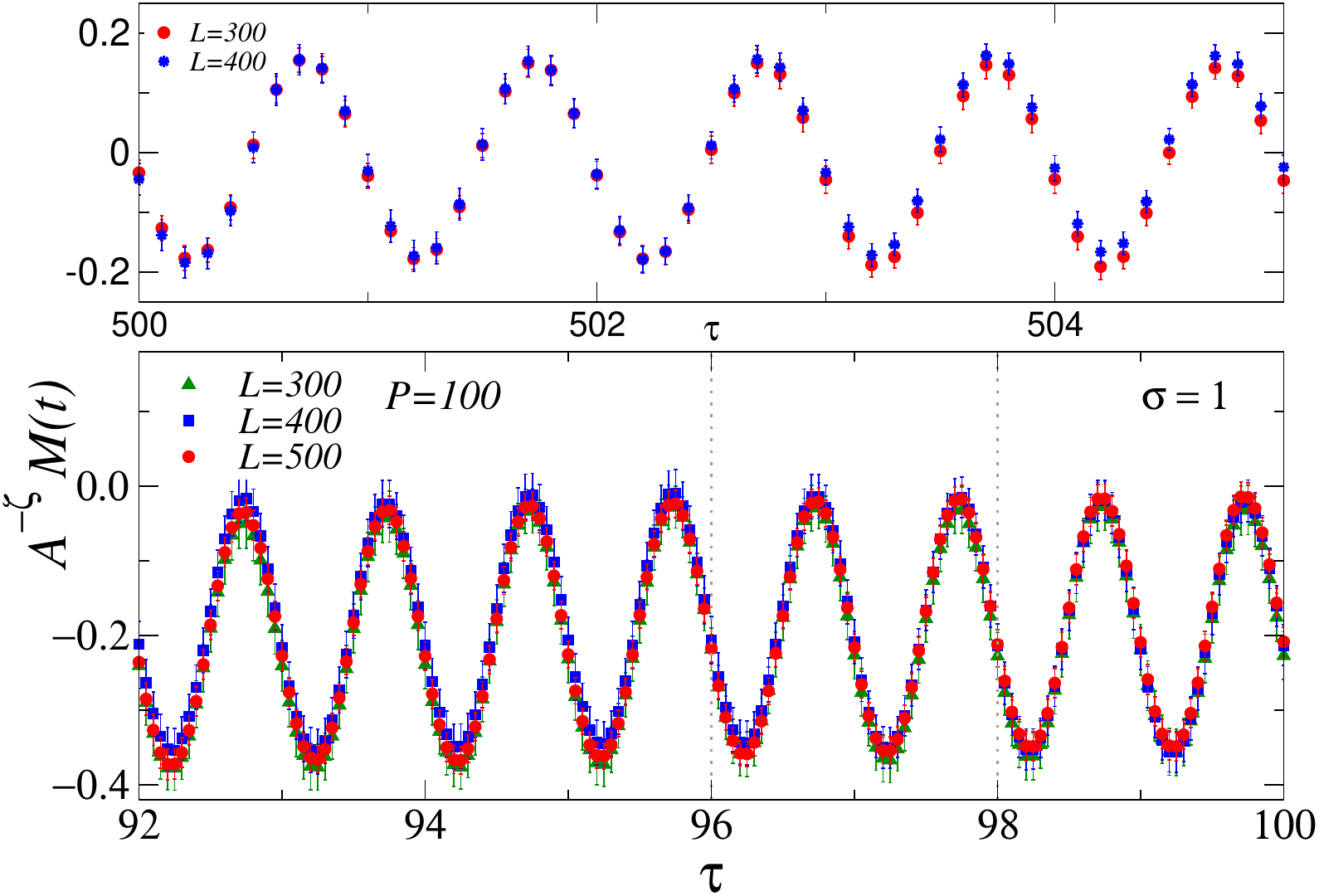}
 \includegraphics*[scale=\graphicscale]{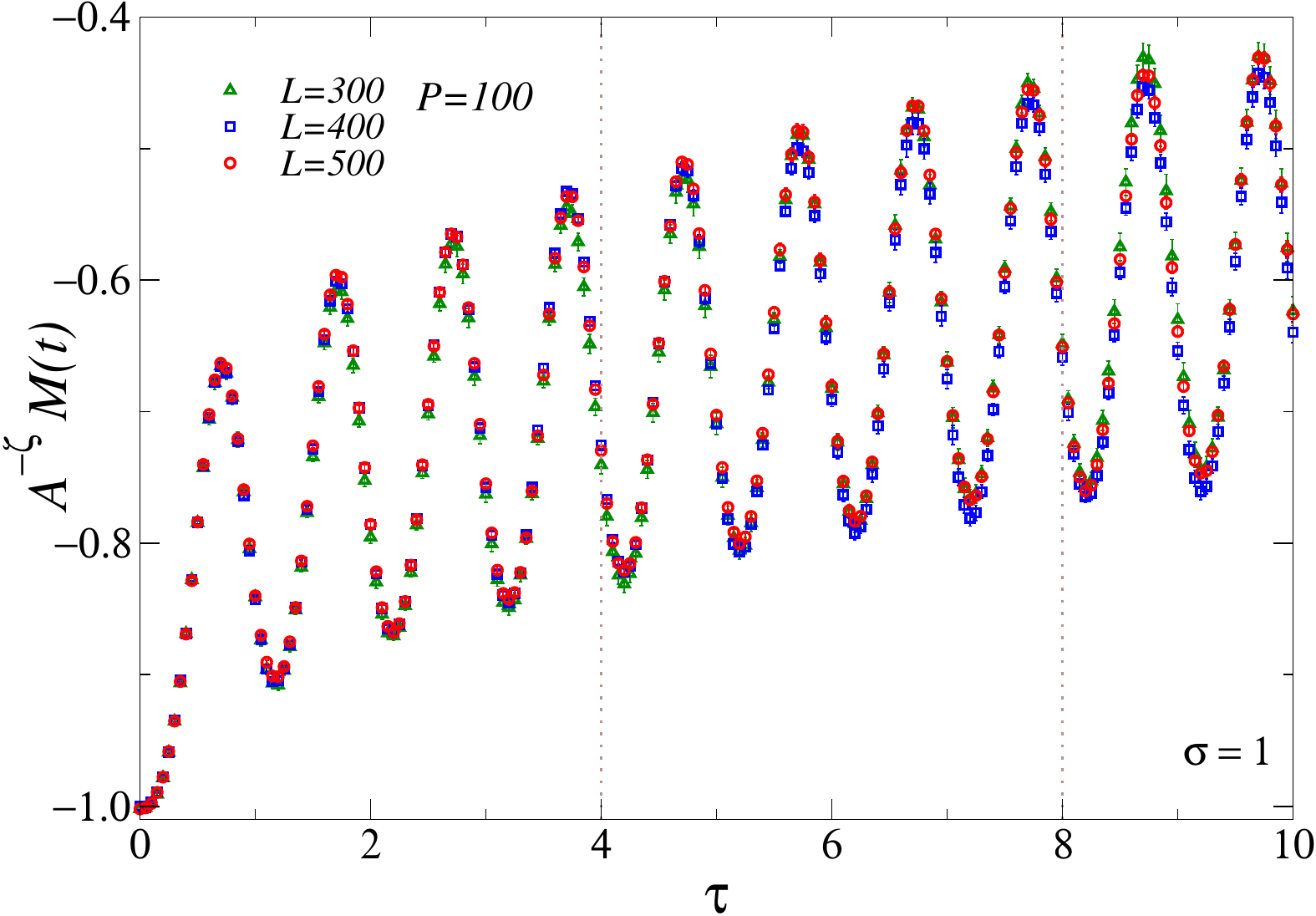}
  \caption{Time evolution of the rescaled magnetization
    $A^{-\zeta}M(t)$ for $P=100$, $\sigma=A P^\kappa=1$, and several
    lattice sizes $L$.  We show results for $\tau \le 10$ (bottom),
    where the center of the oscillations clearly increases, for $92
    \le \tau \le 100$ (middle), where the center of the oscillations
    is still barely increasing, and finally for $500\le \tau \le 505$
    (top), where the behavior is stationary, characterized by a
    vanishing average of the magnetization over one period.  Note that
    the oscillations are synchronized with the external field---their
    period is one in the variable $\tau$---and that the amplitude
    ${\cal A}_m$ does not apparently vary with $\tau$.  In the two
    lower panel we also added dotted vertical lines at integer values
    of $\tau$, corresponding to minima of the magnetic field, to
    highlight the phase delay of the oscillations of the
    magnetization.}
      \label{r1p100}
\end{figure}
%%%%%%%%%%%%%%%%%%%%%%%%%%%%%%%%%%%%%%%%%%%%%%%%%%%%%%%%%%%%%%%%%%%%%%%%

We begin by discussing in detail the behavior for $\sigma = 1$, to
verify the scaling predictions presented in Sec.~\ref{dynsca}. Results
for other values of $\sigma$, which show the same qualitative
behavior, will be discussed later.  Figure~\ref{r1p100} shows results
for the magnetization obtained from protocols with period $P=100$ and
several values of $L\ge 300$.  First, we note that the time-dependent
magnetization approaches the thermodynamic limit with increasing
$L$. Indeed, all data fall onto a unique curve within statistical
errors, showing that the infinite-volume behavior can be accurately
determined by considering systems of size $L\gtrsim 300$.  The time
behavior of the magnetization is characterized by two different
regimes: a small-$\tau$ regime in which the center of the oscillations
is negative and increases with time, and a large-$\tau$ stationary
regime, characterized by symmetric oscillations around $M=0$.  This is
quite evident from the results reported in in Fig.~\ref{r1p100}. In
the two lower panels, in which we show $M(t)$ for $\tau \le 10$ and
$\tau \approx 100$ the magnetization is negative, with an upward trend
that is better visible in the bottom panel. Instead, for $\tau \approx
500$ the stationary regime has been reached, with symmetric
oscillations around zero.

\begin{figure}[!t]
 \includegraphics*[scale=\graphicscale]{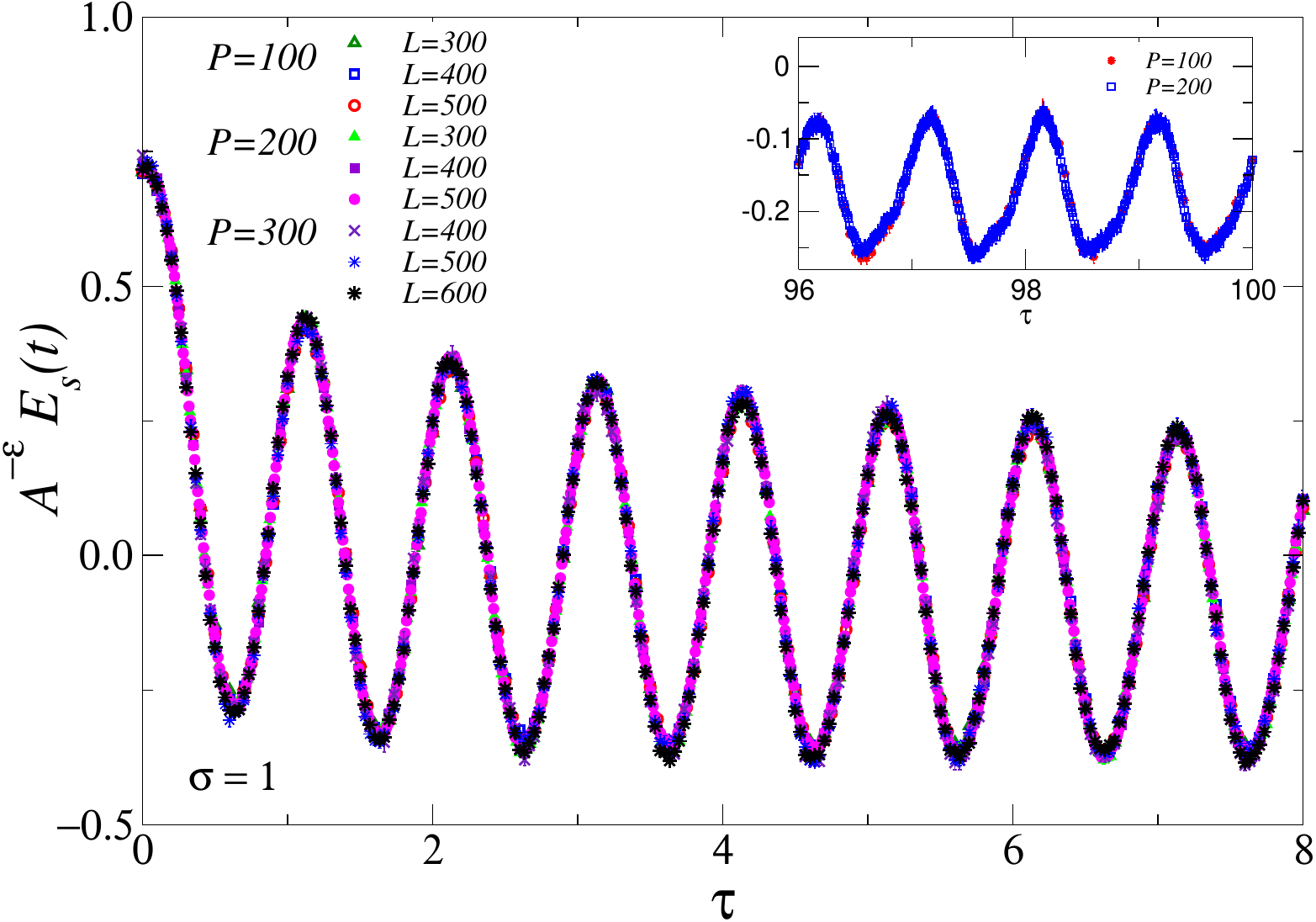}
  \includegraphics*[scale=\graphicscale]{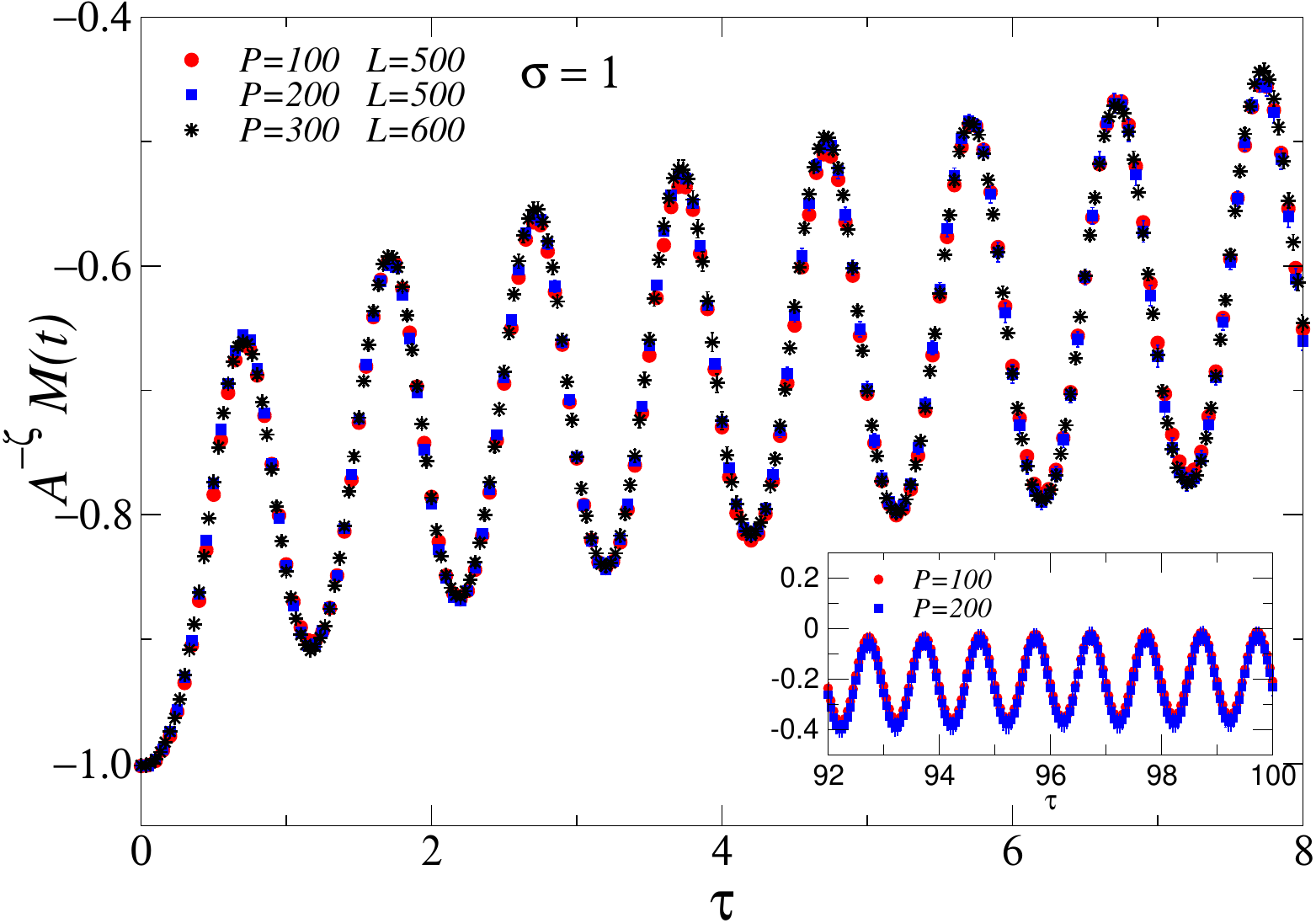}
  \caption{Dynamic scaling behavior of the magnetization (bottom) and
    of the subtracted bond-energy density (top), for $\sigma=1$ and
    several values of $P$, and for $\tau\le 8$. The size $L$ is large
    enough that the data can be considered in the thermodynamic limit.
    In the insets we show the same quantities in time intervals around
    $\tau=100$ (using the data for the largest available lattices).}
      \label{scar1}
\end{figure}

To check the dynamic scaling predictions, we have computed the
magnetization and subtracted bond energy density also for
$P=200,\,300$, always fixing $A$ so that $\sigma=1$, i.e., $A= \sigma
P^{-\kappa}$. We consider $L\gtrsim 300$, which is enough to obtain
results that can be considered in the thermodynamic limit. In
Fig.~\ref{scar1} we plot the rescaled magnetization and subtracted
energy density for different values of $P$: they collapse onto single
curves, fully supporting the expected scaling predictions,
Eq.~(\ref{mtthlim}) and (\ref{enethlim}).  Corrections to the
asymptotic dynamic scaling behavior are smaller than the statistical
errors and not visible on the scale of the figure.

%%%%%%%%%%%%%%%%%%%%%%%%%%%%%%%%%%%%%%%%%%%%%%%%%%%%%%%%%%%%%%%%%%%%%%%%
%%%%%%%%%%%%%%%%%%%%%%%%%%%%%%%%%%%%%%%%%%%%%%%%%%%%%%%%%%%%%%%%%%%%%%%%

%%%%%%%%%%%%%%%%%%%%%%%%%%%%%%%%%%%%%%%%%%%%%%%%%%%%%%%%%%%%%%%%%%%%%%%%
\begin{figure}[!t]
  \includegraphics*[scale=\graphicscale]{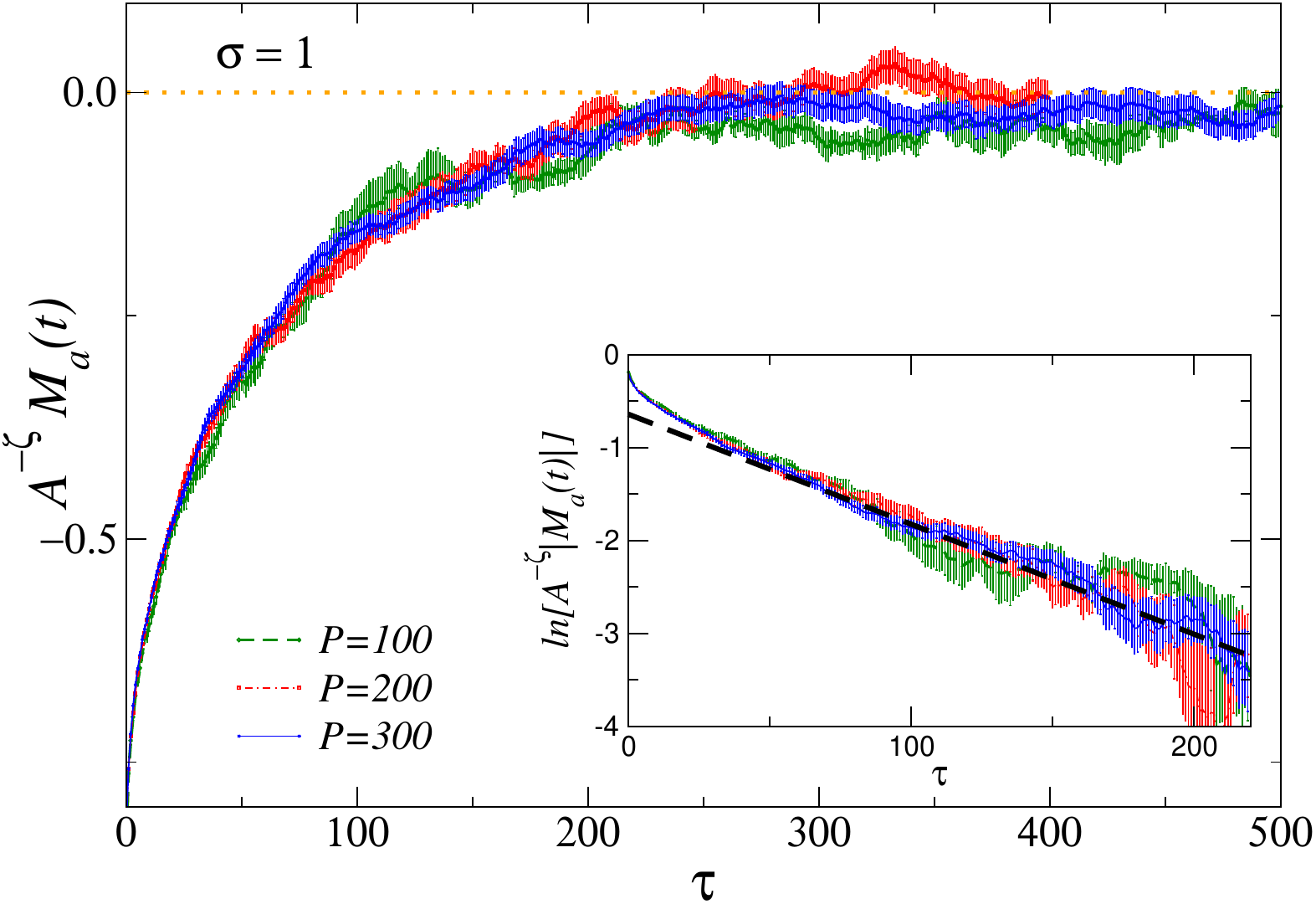}
  \caption{ Period-averaged rescaled magnetization $A^{-\zeta}M_a(t)$,
    for $\sigma=1$, $P=100,\,200,\,300$, in the thermodynamic
    limit. In the inset we use a semilogarithmic scale, to highlight
    the exponential approach to zero: the straight dashed line
    corresponds to a behavior proportional to $\exp(-\tau/\tau_s)$ with
    $\tau_s = 85$.  }
      \label{r1av}
\end{figure}
%%%%%%%%%%%%%%%%%%%%%%%%%%%%%%%%%%%%%%%%%%%%%%%%%%%%%%%%%%%%%%%%%%%%%%%%

\begin{figure}[!t]
      \includegraphics*[scale=\graphicscale]{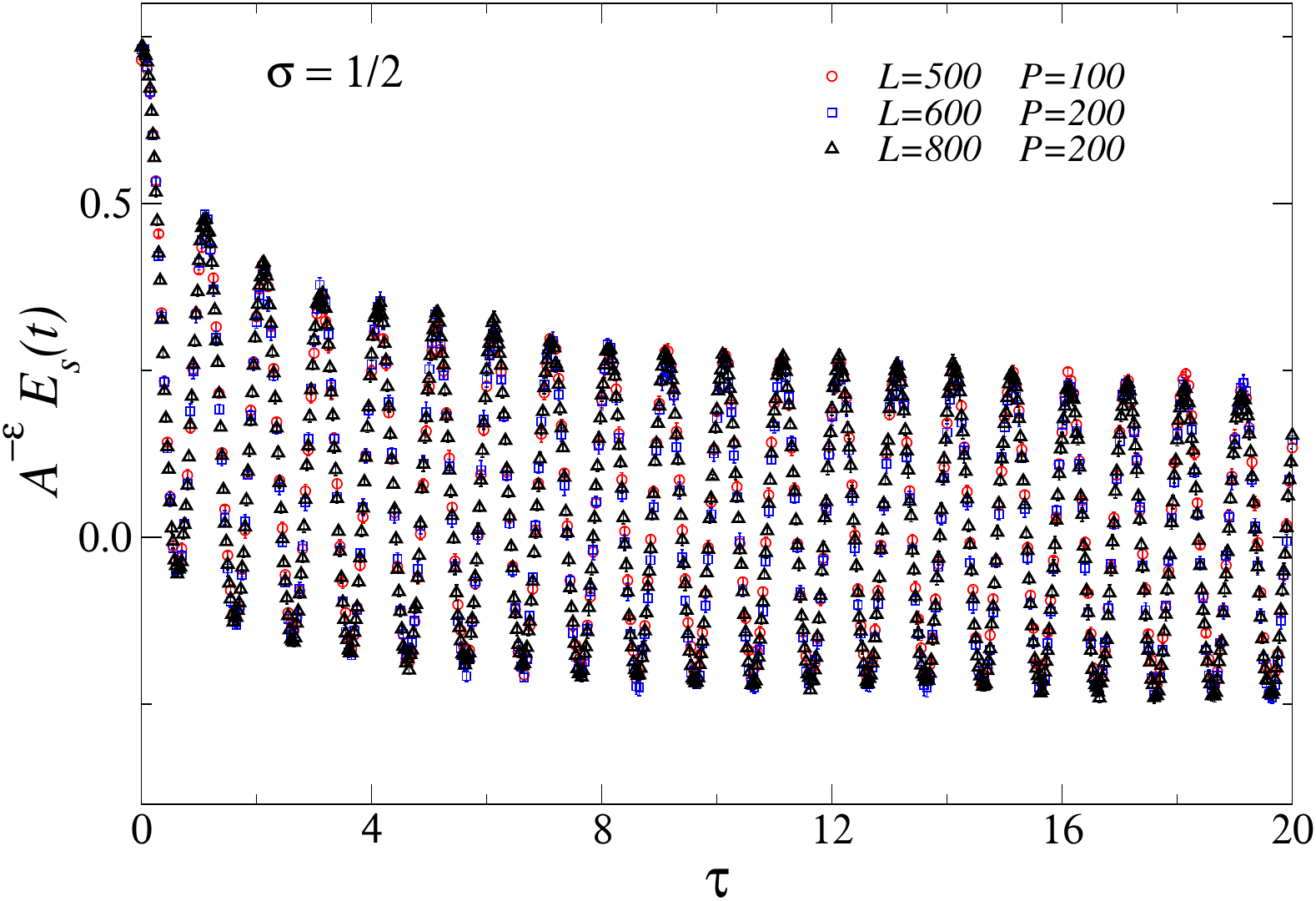}
    \includegraphics*[scale=\graphicscale]{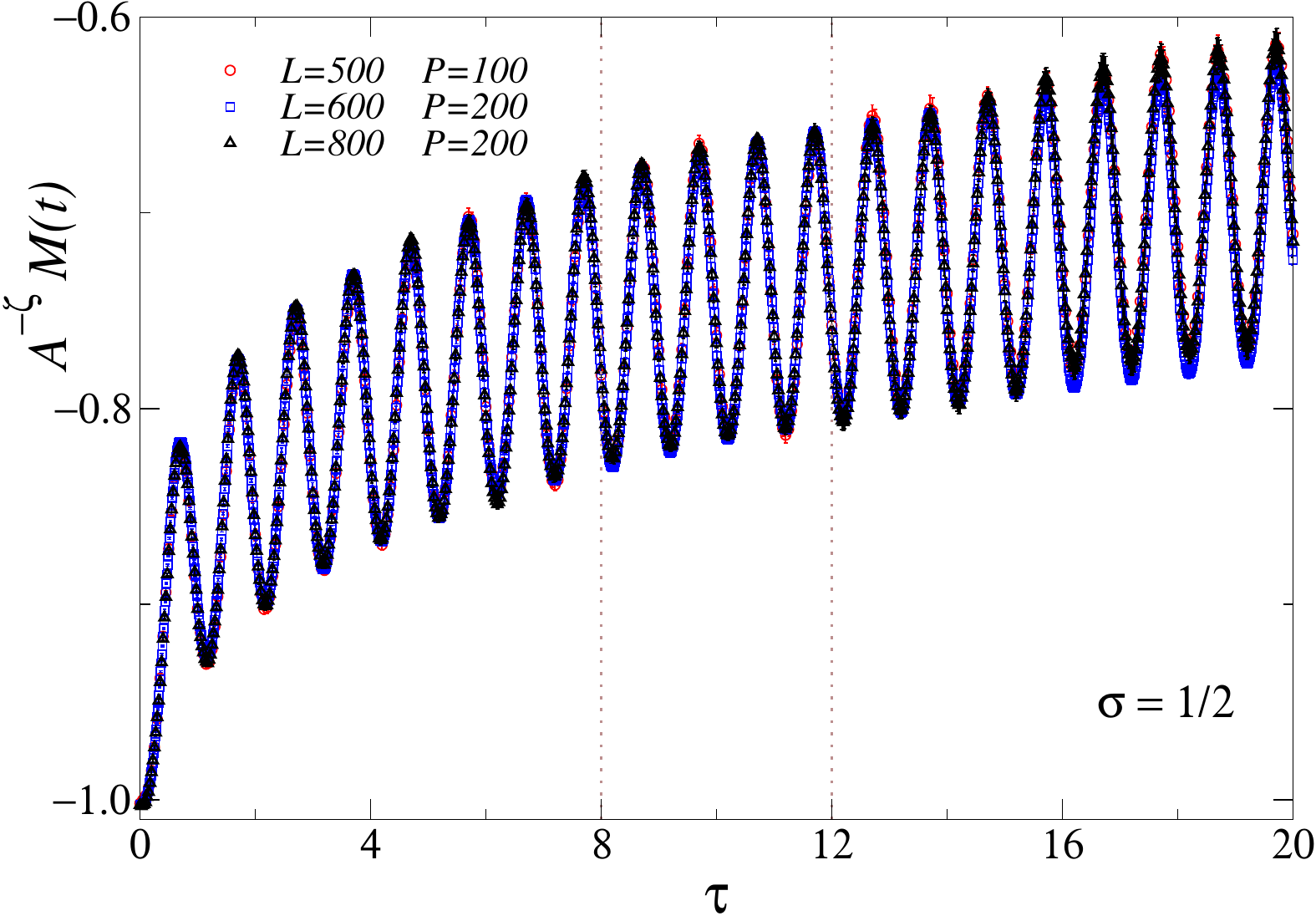}
  \caption{Dynamic scaling results for the magnetization (bottom) and
    for the bond-energy density (top) at fixed $\sigma=A
    P^\kappa=1/2$.  In the lower panel we draw vertical dotted lines
    at some integer values of $\tau$, corresponding to minima of the
    magnetic field, to highlight the phase delay of the oscillations
    of the magnetization.  }
      \label{sig1o2}
\end{figure}

The results shown in Figs.~\ref{r1p100} and \ref{scar1} present a
number of interesting features.  Both the energy density and
magnetization oscillate with period one in $\tau$, showing that the
system is fully synchronized with the magnetic field. The oscillations
however are not in phase with the magnetic field. Indeed, the location
of the minima are shifted by $\delta\tau$ with respect to the integer
values where $h(t)$ has a minimum, see Figs.~\ref{r1p100} and
\ref{sig1o2}.  Numerically, we observe $\delta\tau\approx 0.2$ for
small $\tau$, and $\delta\tau\approx 0.25$ in the stationary regime,
where the minimum is reached close to the point where the magnetic
field changes sign, becoming positive.  The scaling functions show a
smooth drift of the central value of the oscillations, up to
$\tau\approx 300$, where a stationary behavior is reached.  This is
clearly shown by Fig.~\ref{r1av}, which shows the rescaled
period-averaged magnetization defined as
\begin{eqnarray}
  && M_a(t,A,P) = {\sum_{t'=1}^{P} M(t+t',A,P)\over P},
  \label{mavedef}\\
&&A^{-\zeta} M_a(t,A,P) \approx
  {\cal M}_a(\sigma,\tau) = \int_\tau^{\tau+1} d\tau'\,{\cal
    M}(\sigma,\tau').
\nonumber
\end{eqnarray}
Since the data of $M(t,A,P)$ are very correlated, we consider the
statistical error of $M(t,A,P)$ as the uncertainty on the average
$M_a(t,A,P)$.  As shown in Fig.~\ref{r1av}, ${\cal
  M}_a(\sigma=1,\tau)$ is consistent with zero only for $\tau\gtrsim
300$, the approach being apparently exponential, see the inset of
Fig.~\ref{r1av}.  A fit of the data to ${\cal M}_a\sim
\exp(-\tau/\tau_s)$ gives $\tau_s\approx 85$ with a relative
uncertainty of about 10\%.  It is also worth noting that, while the
behavior of ${\cal M}_a(\sigma,\tau)$ becomes stationary only after a
significantly large number of oscillations, the (half) size ${\cal
  A}_m$ of the oscillations (the half difference between successive
maxima and minima) is approximately constant during the time
evolution, i.e., ${\cal A}_m\approx 0.17$ for any $\tau$, see
Figs.~\ref{r1p100} and \ref{scar1}. Similar considerations apply to
the subtracted bond-energy density, which approaches a stationary
behavior for large $\tau$ as well. In this case, oscillations
apparently decrease as $\tau$ increases.

%%%%%%%%%%%%%%%%%%%%%%%%%%%%%%%%%%%%%%%%%%%%%%%%%%%%%%%%%%%%%%%%%%%%%%%%
%%%%%%%%%%%%%%%%%%%%%%%%%%%%%%%%%%%%%%%%%%%%%%%%%%%%%%%%%%%%%%%%%%%%%%%%

%%%%%%%%%%%%%%%%%%%%%%%%%%%%%%%%%%%%%%%%%%%%%%%%%%%%%%%%%%%%%%%%%%%%%%%%
\begin{figure}[!t]
  \includegraphics*[scale=\graphicscale]{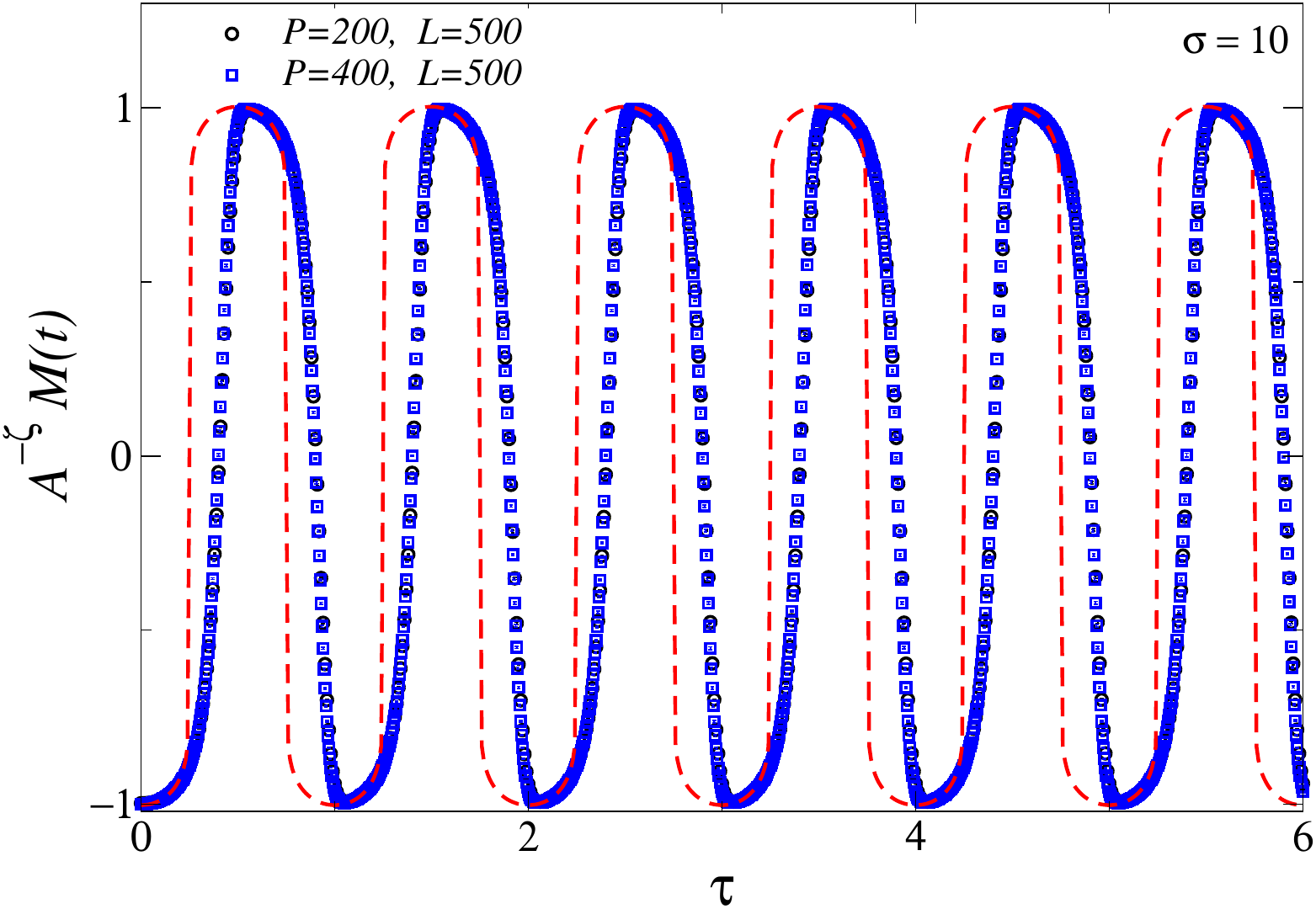}
  \caption{Rescaled magnetization for $\sigma=10$ and different values
    of $P$. Data are obtained on large systems, so they provide an
    accurate approximation of the magnetization in the thermodynamic
    limit. The dashed line is the expected scaling curve in the limit
    $\sigma\to \infty$ reported in Eq.~(\ref{sigmainfi2}).}
      \label{sig10}
\end{figure}
%%%%%%%%%%%%%%%%%%%%%%%%%%%%%%%%%%%%%%%%%%%%%%%%%%%%%%%%%%%%%%%%%%%%%%%%

Analogous results are obtained for $\sigma=1/2,\,3/4,\,2,\,10$.
Results for $\sigma=1/2$ and $\sigma=10$ are shown in
Figs.~\ref{sig1o2} and \ref{sig10}, respectively. We have again
checked that the lattice sizes are sufficiently large to provide an
accurate approximation of the thermodynamic limit. The data confirm
the dynamic scaling predictions reported in Eqs.~(\ref{mtthlim}) and
(\ref{enethlim}). The corresponding scaling curves show features
analogous to those observed for $\sigma=1$.  In particular, the
oscillations are synchronized with the magnetic field for any
$\sigma$, with an amplitude ${\cal A}_m$ that decreases with
decreasing $\sigma$: ${\cal A}_m\approx 0.12$ for $\sigma=3/4$, and
${\cal A}_m\approx 0.08$ for $\sigma=1/2$ (roughly, ${\cal A}_m\sim
\sigma$ for $\sigma\lesssim 1$). For large $\sigma$, ${\cal A}_m$
approaches $c\approx 1$, in agreement with Eq.~(\ref{sigmainfi2}).
The time shift of the minima of the magnetization with respect to the
minima of $h(t)$ is always close to $P/4$ for
$\sigma=1/2,\,3/4,\,1,\,2$.  For $\sigma = 10$ it is significantly
smaller, in agreement with the large-$\sigma$ prediction
(\ref{sigmainfi2}), which implies $\delta\tau\to 0$ for $\sigma\to
\infty$.

%%%%%%%%%%%%%%%%%%%%%%%%%%%%%%%%%%%%%%%%%%%%%%%%%%%%%%%%%%%%%%%%%%%%%%%%
\begin{figure}[!t]
  \includegraphics*[scale=\graphicscale]{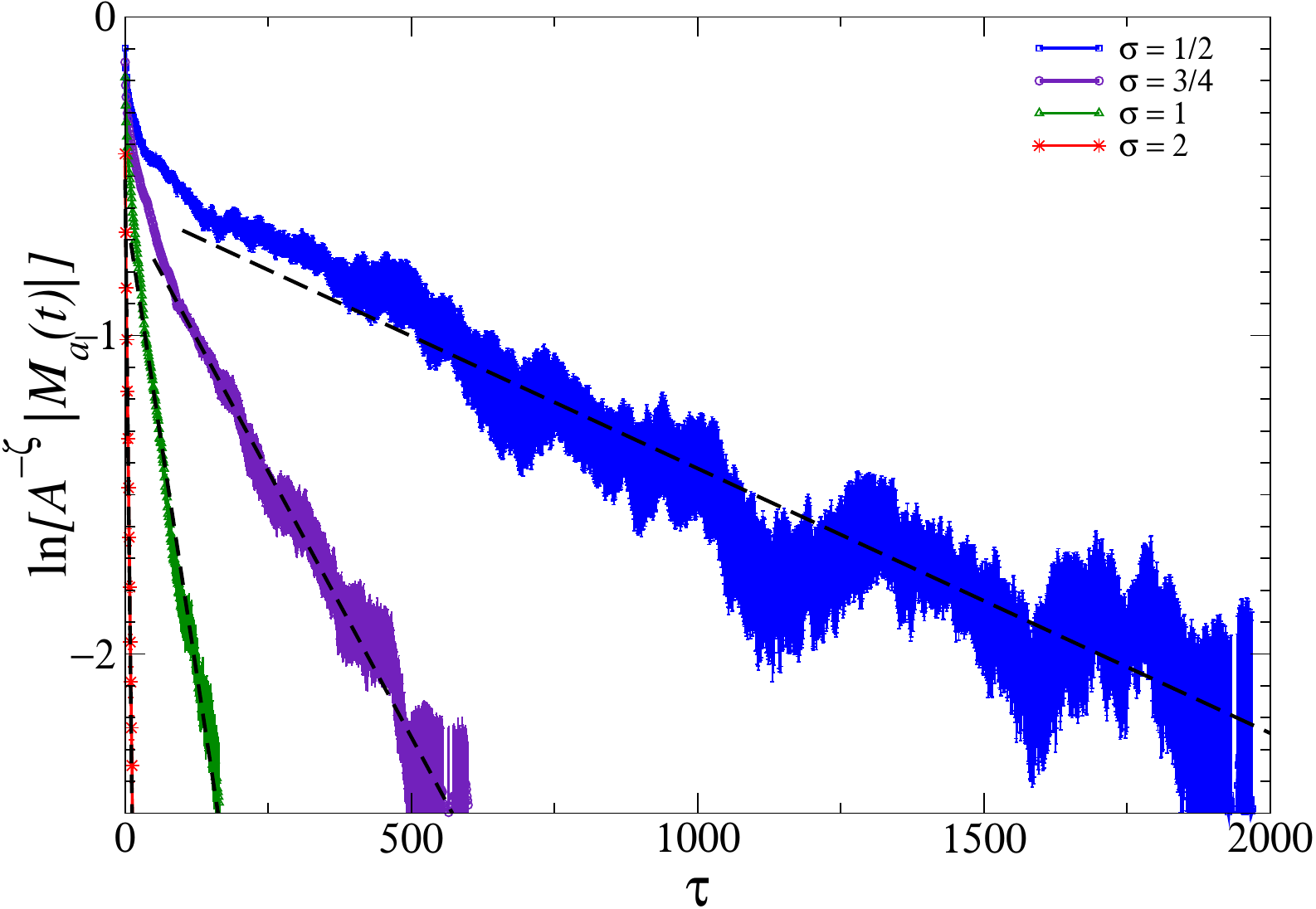}
  \caption{Period-averaged rescaled magnetization for
    several  values of $\sigma$, in the thermodynamic limit.  The
    dashed straight lines correspond to the exponential functions 
    ${\cal M}_a = a e^{-t/\tau_s}$ that interpolate the data: we use
    $\tau_s=6,\,85,\,300,\,1200$ for
    $\sigma=2,\,1,\,3/4,\,1/2$, respectively.}
      \label{ravl}
\end{figure}
%%%%%%%%%%%%%%%%%%%%%%%%%%%%%%%%%%%%%%%%%%%%%%%%%%%%%%%%%%%%%%%%%%%%%%%%

For $\sigma = 10$, the magnetization ${\cal M}(\sigma,\tau)$ reported
in Fig.~\ref{sig10} always shows a stationary behavior, with
oscillations that are symmetric around zero. The corresponding curve
is also close to the one expected in the $\sigma\to\infty$ limit,
reported in Eq.~(\ref{sigmainfi2}). On the other hand, for
$\sigma\lesssim 2$, the stationary behavior sets in at finite values
of $\tau$ as already discussed for $\sigma=1$: the period-averaged
magnetization ${\cal M}_a(\sigma,\tau)$ increases towards zero with a
time scale $\tau_s(\sigma)$ that strongly increases with decreasing
$\sigma$; see Fig.~\ref{ravl}, where the magnetization data are
reported for various values of $\sigma$. We conjecture that ${\cal
  M}_a(\sigma,\tau)$ converges to zero for any, however small, value
of $\sigma>0$, so in all cases the evolution recovers the ${\mathbb
  Z}_2$ parity invariance broken by the starting conditions.  However,
while we have a robust numerical evidence that the period-averaged
magnetization vanishes in the large-$\tau$ limit for $\sigma\ge 1$,
for $\sigma<1$ we observe that the absolute value of the magnetization
decreases, but we are not able to clearly identify the stationary
regime.  The data shown in Fig.~\ref{ravl}, are consistent with the
asymptotic exponential behavior
\begin{equation}
{\cal M}_a(\sigma,\tau) \sim e^{-\tau/\tau_s(\sigma)}, 
\label{maexpasy}
\end{equation}
for sufficiently large values of $\tau$.  The time scale
$\tau_s(\sigma)$ rapidly increases with decreasing $\sigma$.  Indeed,
we estimate $\tau_s\approx 6$ for $\sigma=2$, $\tau_s\approx 85$ for
$\sigma=1$, $\tau_s\approx 300$ for $\sigma=3/4$, and $\tau_s\approx
1200$ for $\sigma=1/2$ (they are obtained by fitting the data in the
range $\ln {\cal M}(\sigma,\tau)\in [-3,-1]$).  Due to the large
fluctuations of the data, the uncertainty on these estimates is
relatively large, about 10\% for $\sigma=1,\,2$, and about 20\% for
$\sigma=3/4,\,1/2$ (we remark that a substantial improvement of the
estimates of $\tau_s$ for small $\sigma$ would require a significant
additional numerical effort).  Although these estimates should be
considered as indicative, it is worth noting that they are consistent
with a power-law behavior for $\sigma\to 0$, i.e., with
\begin{equation}
  \tau_s(\sigma) \approx  a_s \, \sigma^{-u}.
  \label{taussig}
\end{equation}
A linear fit of the above estimates of $\tau_s(\sigma)$ to
Eq.~(\ref{taussig}) gives the estimate $u=3.9(2)$.  Note that the
exponent $u$ turns out not to be related to the standard critical
exponents of the Ising transition.  The divergence of $\tau_s(\sigma)$
indicates that, if we take the limit $\sigma\to 0$ at fixed $\tau$, we
always obtain ${\cal M}(\sigma\to 0,\tau)\approx {\cal M}(0,0)$, in
agreement with Eq.~(\ref{sigma02}).

\subsection{Single-system time behavior}

%%%%%%%%%%%%%%%%%%%%%%%%%%%%%%%%%%%%%%%%%%%%%%%%%%%%%%%%%%%%%%%%%%%%%%%%
\begin{figure}[!t]
 \includegraphics*[width=8.4cm]{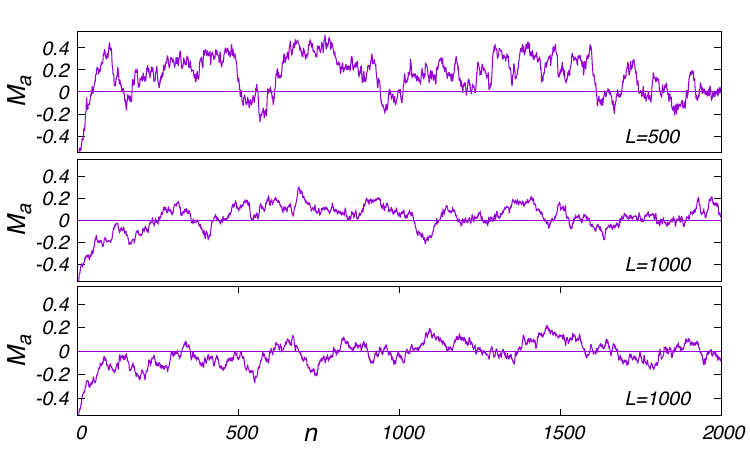}
 \includegraphics*[width=8.4cm]{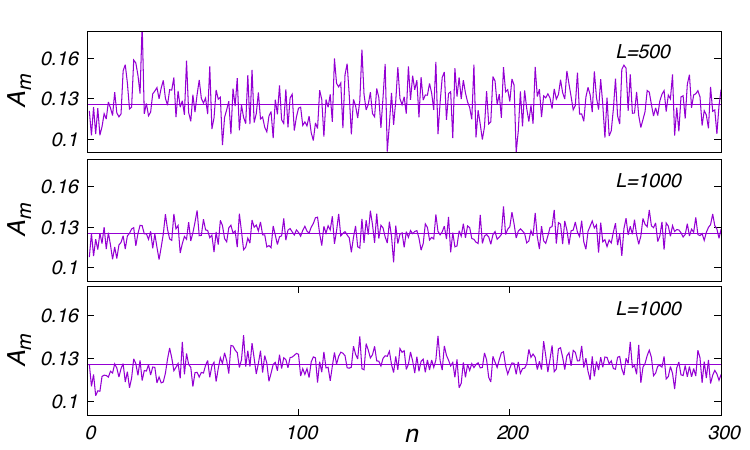}
\caption{Time evolution of the period-averaged magnetization $M_a(n)$
  (upper three panels) and of the amplitude $A_m(n)$ of the
  oscillations of the magnetization in a period (lower three panels)
  for three different evolutions, as a function of the number of
  periods $n$.  Here $P=200$ and $\sigma = 1$.  Two evolutions refer
  to systems of size $L=1000$, one evolution to a system of size
  $L=500$.  The straight horizontal line in the amplitude plots
  corresponds to $A_m = 0.1254$.  Note we use a different horizontal
  scale for $M_a$ and $A_m$. Instead, we use the same vertical scale
  for data corresponding to systems of different size.  }
\label{single-evolution}
\end{figure}
%%%%%%%%%%%%%%%%%%%%%%%%%%%%%%%%%%%%%%%%%%%%%%%%%%%%%%%%%%%%%%%%%%%%%%%%

%%%%%%%%%%%%%%%%%%%%%%%%%%%%%%%%%%%%%%%%%%%%%%%%%%%%%%%%%%%%%%%%%%%%%%%%
\begin{figure}[!t]
 \includegraphics*[width=8cm]{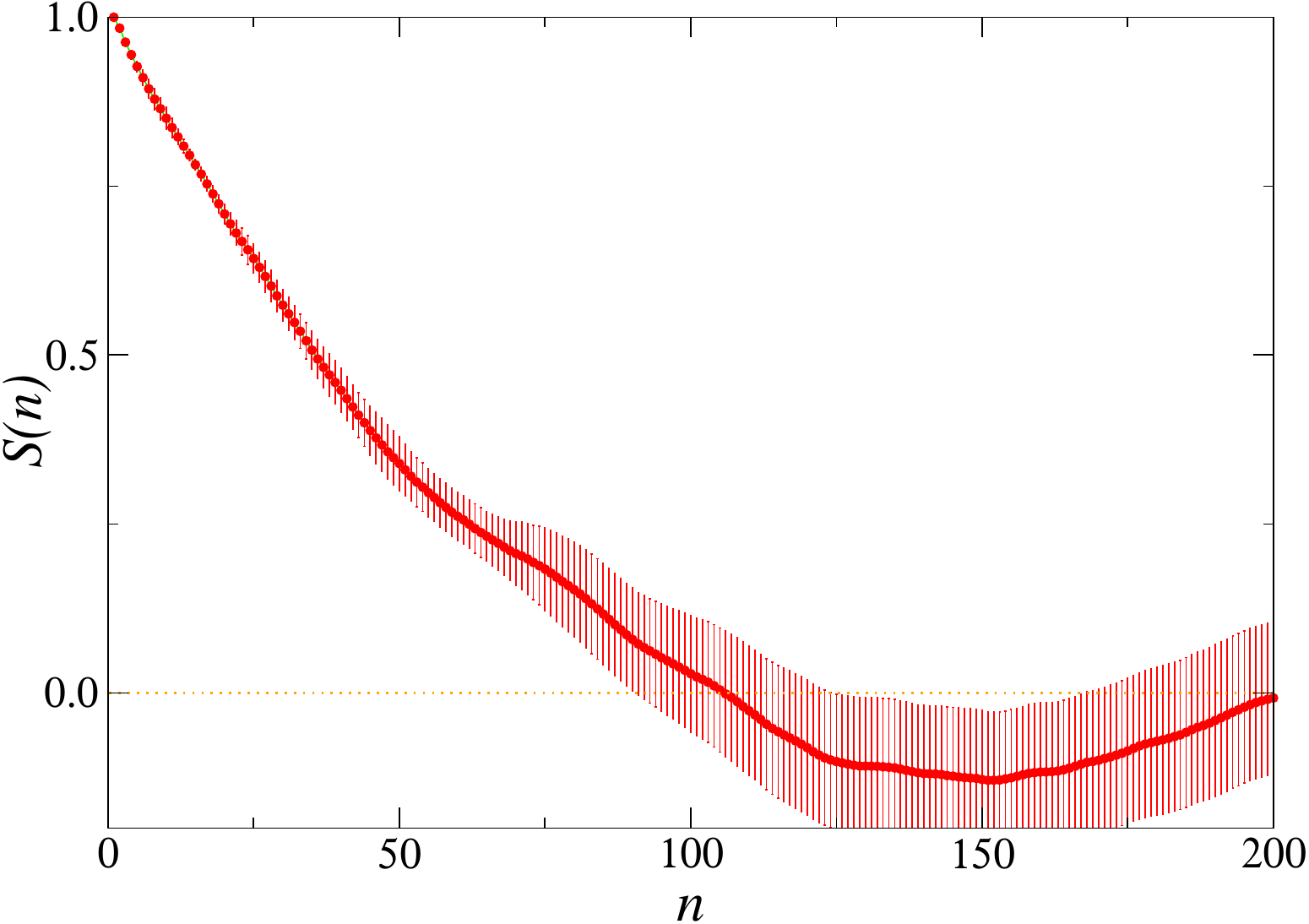}
 \caption{Normalized autocorrelation function $S(n)$ for the
   period-averaged magnetization $M_a(n)$ for $L=1000$, $P=200$,
   $\sigma = 1$.}
\label{autocorrf}
\end{figure}
%%%%%%%%%%%%%%%%%%%%%%%%%%%%%%%%%%%%%%%%%%%%%%%%%%%%%%%%%%%%%%%%%%%%%%%%

In the previous section we have performed a detailed study of the
average behavior of the system as a function of time, the average
being taken over different time trajectories and starting
configurations. In this section we focus on the evolution of a a
single system as a function of time.  We consider $P = 200$, fix the
amplitude $A$ of the magnetic field so that $\sigma = 1$, i.e.,
$A=\sigma P^{-\kappa}\approx 0.0102$, and study the behavior of the
period-averaged magnetization $M_a(n)$, defined in
Eq.~(\ref{mavedef}), $n$ being an integer number.\footnote{We use here
the same notations as in the previous section, but note that in this
section $M(t)$ and $M_a(n)$ refer to a single system and there is no
average over trajectories.}  We have also studied the evolution for
$\sigma = 2$, obtaining qualitatively similar results.

The average $M_a(n)$ is reported in the top panels of
Fig.~\ref{single-evolution}, for two different values of $L$. In all
cases it undergoes large oscillations with a period of the order of
$100P$.  Moreover, as in the average case, see Fig.~\ref{r1av}, for
small values of $n$ the magnetization is negative; a stationary
behavior, characterized by fluctuations around zero, is only reached
after a transient region also of the order of $100 P$.  These results
are in agreement with those reported before, that showed that the
average magnetization $M_a(t)$ behaves as $M_a(t) \sim
\exp(-\tau/\tau_s)$, with $\tau_s\approx 85$. To characterize the
fluctuations in the stationary regime, we compute the autocorrelation
function
\begin{eqnarray}
&& S(n) =  {C_M(n) \over C_M(0)}, \label{autocorr}\\
&& C_M(n)  = 
     {\sum_{m = n_{\min}+1}^{n_{\max}}  M_a(m) M_a(m + n)\over
n_{\max} - n_{\min} - n}. \nonumber 
\end{eqnarray}
Results are reported in Fig.~\ref{autocorrf}, for $L=1000$, $n_{\rm
  min} = 500$, and $n_{\rm max} = 2000$. They clearly indicate that
also in the stationary state there are strong time correlations of the
order of $100P$. We have not been able to reliably estimate an
autocorrelation time from the data, as results strongly depend on the
chosen fitting form. If we fit $S(n)$ to $Q(n) e^{-n/\tau_s}$, where
$Q(n)$ is a polynomial in $n$, we obtain estimates of $\tau_s$ that
vary between 60 and 120, depending on the order of the polynomial. In
any case, they are fully compatible with the estimate $\tau_s \approx
85$ obtained from the analysis of the small-$\tau$ transient behavior
presented in the previous section.

%%%%%%%%%%%%%%%%%%%%%%%%%%%%%%%%%%%%%%%%%%%%%%%%%%%%%%%%%%%%%%%%%%%%%%%%
\begin{figure}[!t]
 \includegraphics*[width=8cm]{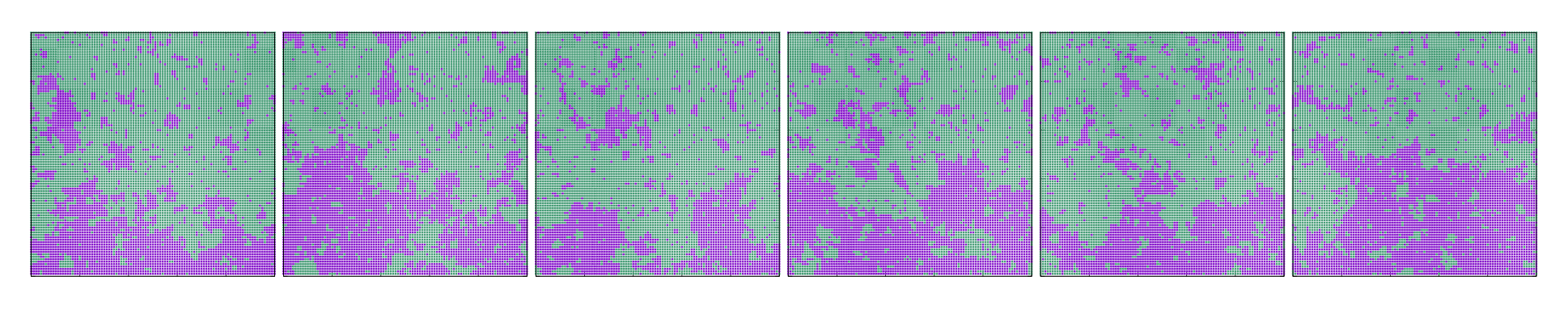}
 \includegraphics*[width=8cm]{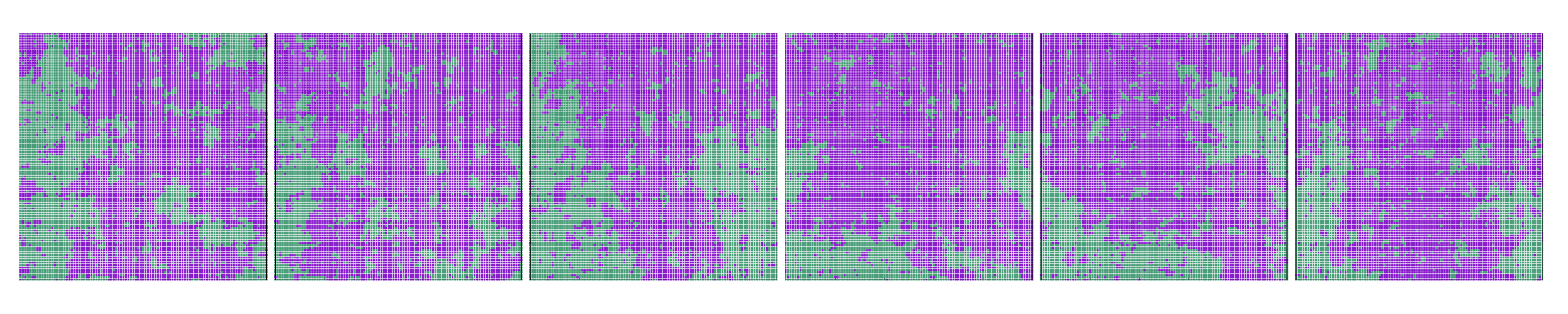}
 \caption{Evolution of the spin configurations of two different
   sublattices of size $100^2$ for a large system of size $L=1000$;
   results for $P=200$ and $\sigma = 1$.  Green sites correspond to
   negative spins, magenta sites to positive spins.  Each horizontal
   line corresponds to a sublattice at time $t_1 = 500P + 50$, $t_2 =
   t_1 + P/2$, $t_1 + P$, $\ldots$, $t_6 = t_1 + 5P/2$ (going from
   left to right time increases by $P/2$ each time).  At times
   $t=t_1$, $t_3$, $t_5$, $M(t)$ is very close to the minimum value of
   the magnetization in the oscillation, $M \approx M_{\rm min}$,
   while at time $t=t_2$, $t_4$, $t_6$, $M(t) \approx M_{\rm max}$.  }
\label{snapshot}
\end{figure}
%%%%%%%%%%%%%%%%%%%%%%%%%%%%%%%%%%%%%%%%%%%%%%%%%%%%%%%%%%%%%%%%%%%%%%%%

The presence of large-time correlations can also be inferred by
directly looking at the evolution of the spin configurations.  In
Fig.~\ref{snapshot} we report the time evolution of two sublattices of
size $100^2$, belonging to a large system of size $L=1000$.  Time
varies going from left to right, the time distance between two
successive panels being $P/2$. One can immediately observe that, in a
time span of three periods, the large-scale nature of the two
sublattices does not change: the sublattice reported in the upper part
of the figure is always negatively magnetized (green points correspond
to $s_{\bm x} = -1$), while the sublattice reported in the lower part of the
figure is always positively magnetized (positive spins are represented
by magenta points). Clearly significant changes only occur over many
oscillations.

The magnetization data reported in Fig.~\ref{single-evolution} also
indicate that the fluctuations of $M_a(t)$ decrease as $L$ increases.
We compute the standard deviation $\sigma_M$ of the fluctuations---it
corresponds to $\sqrt{C_M(0)}$, where $C_M(n)$ is defined in
Eq.~(\ref{autocorr})---obtaining $\sigma_M \approx 0.162$ for the run
with $L=500$ and $\sigma_M \approx 0.0733$ and $0.0844$ for the two
runs with $L=1000$.  Fluctuations decrease as $1/L$, as
expected. Indeed, $\sigma_M^2 = \chi L^{-2}$, where $\chi$ is the
time-dependent susceptibility. If we take the limit $L\to \infty$ at
fixed Hamiltonian parameters, $\chi$ converges to its infinite-volume
finite value, so $\sigma_M^2 \sim L^{-2}$, as confirmed by the
data. Since $\sigma_M \to 0$ for $L\to\infty$, the average
magnetization satisfies self-averaging: if $L$ is very large, the time
behavior of $M_a$ for a single system is essentially the same as the
average behavior computed in the previous section. As we shall see
below, also the amplitude of the oscillations becomes
system independent for $L\to \infty$, indicating that the
magnetization $M(t)$ itself satisfies self-averaging.

We have also analyzed the time and size behavior of the amplitude of 
the oscillations. We define 
\begin{eqnarray}
  && A_m(n) = {1\over2} \bigl[M_{\rm max}(n) - M_{\rm min}(n)\bigr],
  \label{amdef}\\
&& M_{\rm max}(n) = \max_{t\in[n P, (n + 1)P]} M(t),  \nonumber\\
&& M_{\rm min}(n) = \min_{t\in[n P, (n + 1)P]} M(t).  \nonumber 
\end{eqnarray}
Its time behavior is reported in the lower panels of
Fig.~\ref{single-evolution}. The behavior of $A_m(n)$ is quite
different from that of $M_a(n)$. Indeed, data are uncorrelated and
there is little evidence of an initial transient behavior.  The
average of the amplitude over time shows very little $L$ dependence
and we estimate $\langle A_m\rangle_n \approx 0.1254$ and 0.1253 for
the two runs with $L=1000$ and 0.1260 for the run with $L=500$. Also
in this case fluctuations decrease as $1/L$. Indeed, the standard
deviation of the fluctuations is equal to 0.00342, 0.00347 for the two
runs with $L=1000$, and $0.00707$ for $L=500$.

\subsection{Behavior under a square-wave protocol}
\label{squarewave}

%%%%%%%%%%%%%%%%%%%%%%%%%%%%%%%%%%%%%%%%%%%%%%%%%%%%%%%%%%%%%%%%%%%%%%%%
\begin{figure}[!t]
 \includegraphics*[scale=\graphicscale]{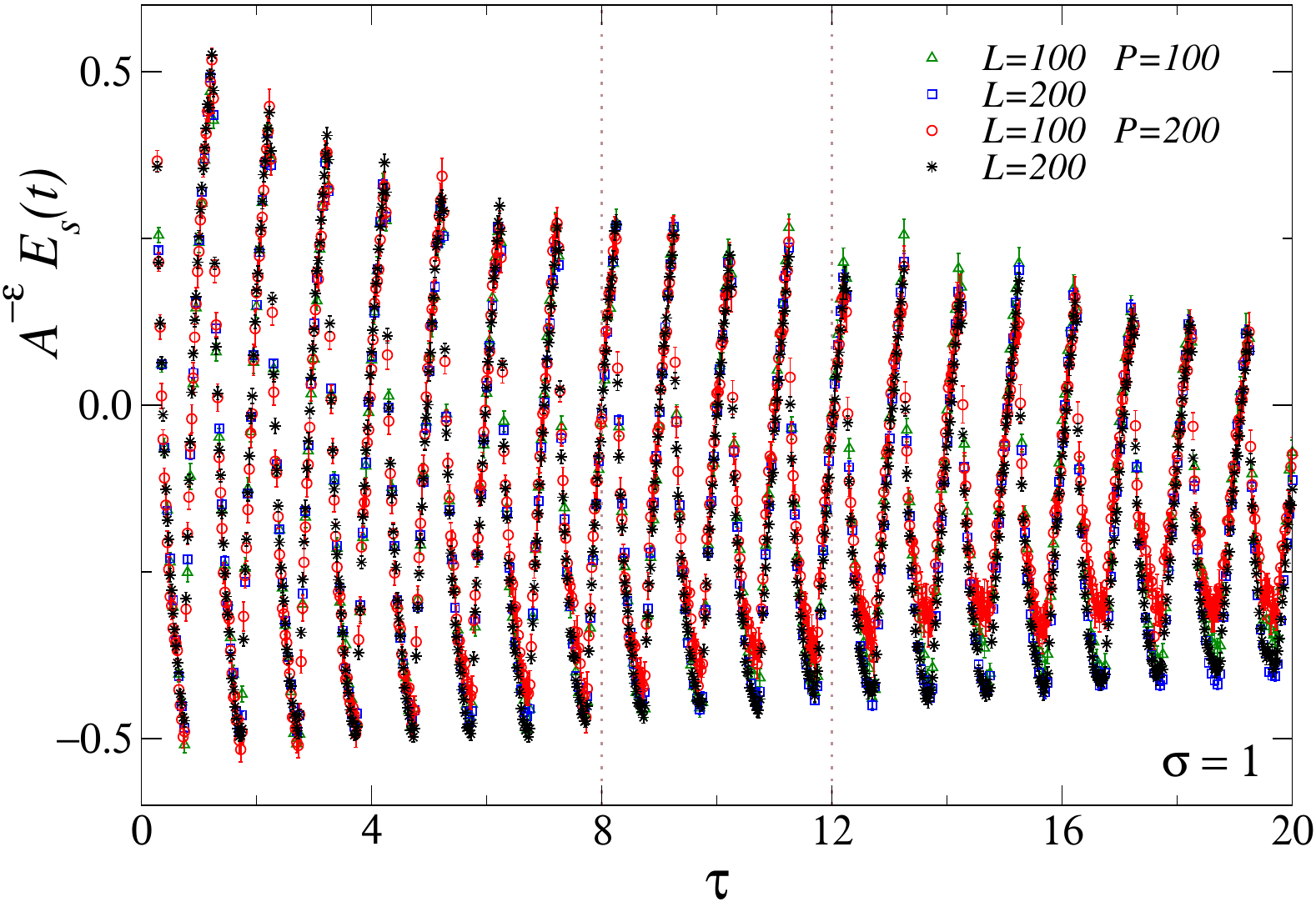}
 \includegraphics*[scale=\graphicscale]{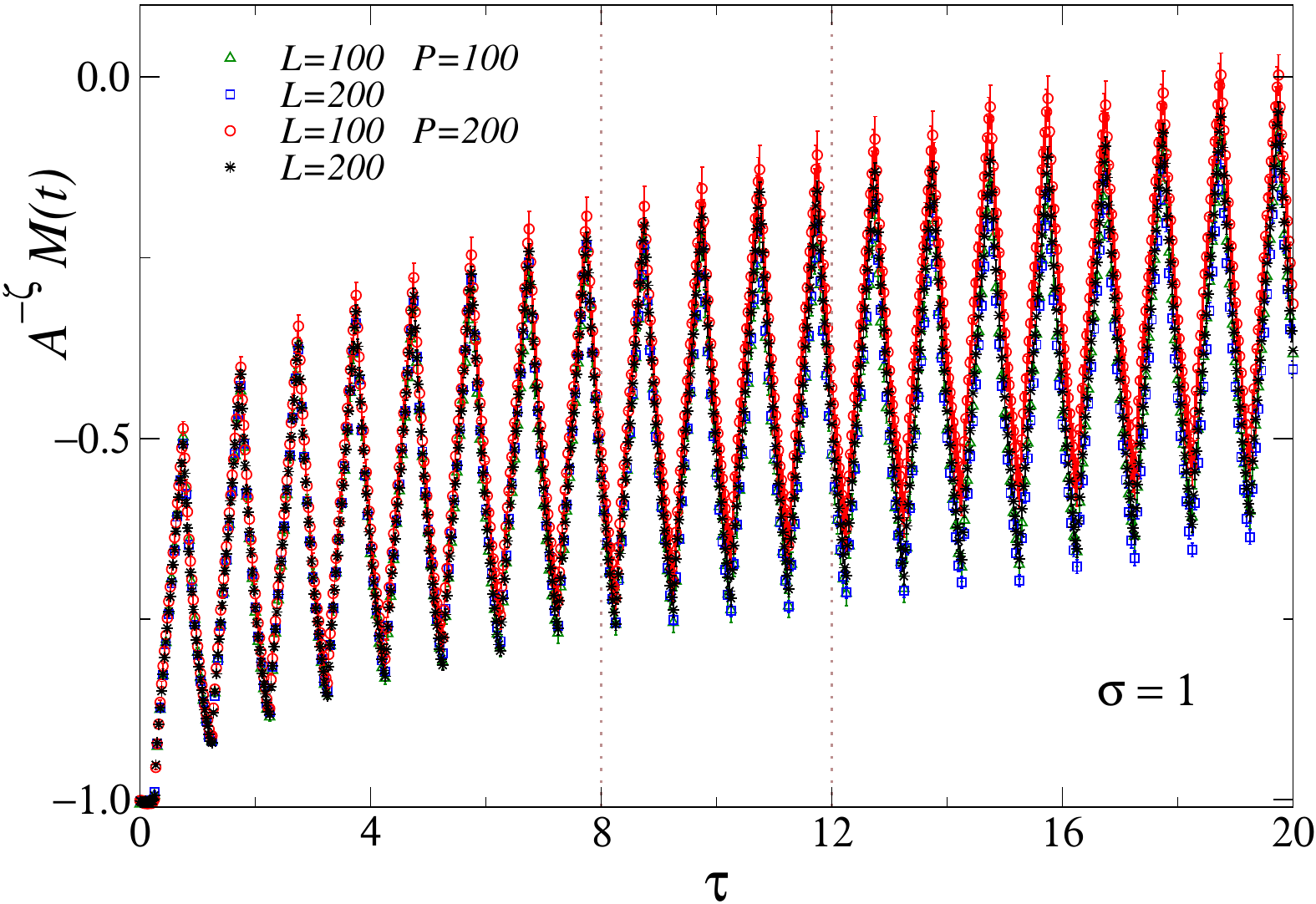}
 \caption{Dynamic scaling of the magnetization (bottom) and of the 
   bond-energy density (top) under a square-wave periodic driving,
   see Eq.~(\ref{hst}).  We report data for $\sigma=A P^\kappa=1$, 
   for two different values of $L$, to verify that we are essentially
   probing the thermodynamic limit, and 
   for two different values of $P$, to verify the 
   the expected dynamic scaling behavior.  The dotted
   lines at some integer values of $\tau$ are reported to highlight
   the phase delay of the oscillations of the magnetization with
   respect to those of the magnetic field $h_s(t)$.  }
      \label{qrchi}
\end{figure}
%%%%%%%%%%%%%%%%%%%%%%%%%%%%%%%%%%%%%%%%%%%%%%%%%%%%%%%%%%%%%%%%%%%%%%%%

To verify that the qualitative features of the observed scaling
behavior are not specific of the protocol we have considered, we now
briefly study the magnetization behavior under a square-wave periodic
driving.  We assume that the magnetic field varies as
\begin{eqnarray}
h_s(t) = - A \; {\rm sgn}\Bigl[\cos(2\pi t/P)\Bigr], \qquad h_s(0) = - A,
\label{hst}
\end{eqnarray}
where ${\rm sgn}(x)$ is the sign function [defined by ${\rm sgn}(x) =
x/|x|$ and ${\rm sgn}(0)=0$].  The dynamic scaling theory in the
thermodynamic limit is analogous to that derived for the protocol with
the cosine dependence (\ref{htdef}).  This is confirmed by the
numerical results shown in Fig.~\ref{qrchi}, which fully support the
dynamic scaling relations, Eqs.~(\ref{mtthlim}) and (\ref{enethlim}).
Although the resulting scaling functions differ quantitatively, they
present features analogous to those displayed in the presence of a
sinusoidal driving, such as synchronization with a phase delay.

\section{Periodic temperature driving at zero magnetic field}
\label{tempdriv}

In this section we analyze a different protocol characterized by a
time-varying temperature at zero magnetic field. More precisely, we
consider the evolution of a system with Hamiltonian (\ref{classisi})
at $h=0$, with inverse temperature varying as
\begin{equation}
  \delta \beta(t) \equiv \beta(t)-\beta_c = - B\,{\rm cos}(2\pi t/P),
  \label{deltabetat}
\end{equation}
with $B > 0$.
The evolution starts from equilibrium configurations at the initial inverse
temperature $\beta_i = \beta_c - B<\beta_c$. We again
study the effects of this periodic driving on the infinite-volume evolution
of some large-scale observables.

Since the magnetization vanishes due to the invariance of the Hamiltonian 
under ${\mathbb Z}_2$ parity transformations,
we consider the time-dependent susceptibility
\begin{equation}
  \chi(t,B,P) =  {1\over L^2} \langle (\sum_{\bm x} s_{\bm x})^2
  \rangle_t\,.
  \label{chitdef}
\end{equation}
As a second observable we consider a quantity related to the energy.
In the present case its definition requires some care. Indeed, in the
equilibrium case, the standard RG analysis shows that, beside the
presence of a mixing with the identity operator there are also
additional terms due to an eigenvalue resonance (the identity and
energy RG dimensions differ by an integer), which give rise to
next-to-leading logarithmic terms~\cite{Wegner-76,PV-02},
\begin{equation}
  E_e(\beta) \approx E(\beta_c) + c\,(\beta-\beta_c)\ln|\beta-\beta_c|+...,
    \label{eebeta}
\end{equation}
These mixings must be subtracted in order to obtain the correct RG
energy eigenoperator, with RG dimension $y_e=d-y_t$, thus $y_e=1$ for
the 2D Ising transition. Thus, a simple subtraction of the critical
bond-energy value $E_c$, as in the case of a magnetic
driving,\footnote{In the case of a magnetic driving, the subtraction
of $E_c$ provides the correct scaling operator, because of a
nontrivial, and as yet unexplained, feature of the scaling functions
that parametrize the logarithmic singularity, see
Refs.~\cite{AF-80,CHPV-02,CGNP-11}: they are independent of the
magnetic field. If this were not the case, one should have performed a
more careful definition of the subtracted bond energy also in the
magnetic case.} does not provide the correct lattice quantity, which
can be identified with the energy RG eigenoperator.

To obtain the correct scaling quantity, we consider a different
subtracted bond-energy density, defined by subtracting the equilibrium
value at the given instantaneous temperature,
\begin{eqnarray}
E_{se}(t,B,P) \equiv E(t) - E_e[\beta(t)].
    \label{esdef2}
\end{eqnarray}
Here $E(t)$ is the bond-energy density defined in Eq.~(\ref{esdeft}),
and $E_e(\beta)$ is the equilibrium bond-energy density at the given
inverse temperature $\beta$,
i.e.,~\cite{Onsager-44,Kaufman-49,Baxter-book}
\begin{eqnarray}
&&E_{e}(\beta) = {\rm coth}(2\beta)
  \Bigl\{ 1 + {2\over \pi}\Bigl[2\,{\rm tanh}(2\beta)^2 - 1\Bigr]
  \,K[k(\beta)]^2\Bigr\},
\nonumber\\
&&k(\beta) = 2{{\rm sinh}(2\beta)\over{\rm cosh}(2\beta)^2}, 
\label{eeui}
\end{eqnarray}
where $K$ is the complete elliptic integral of the first kind.  One
can easily check that $E_e(\beta_c) = \sqrt{2}$.

%%%%%%%%%%%%%%%%%%%%%%%%%%%%%%%%%%%%%%%%%%%%%%%%%%%%%%%%%%%%%%%%%%%%%%%%
\begin{figure}[!t]
 \includegraphics*[scale=\graphicscale]{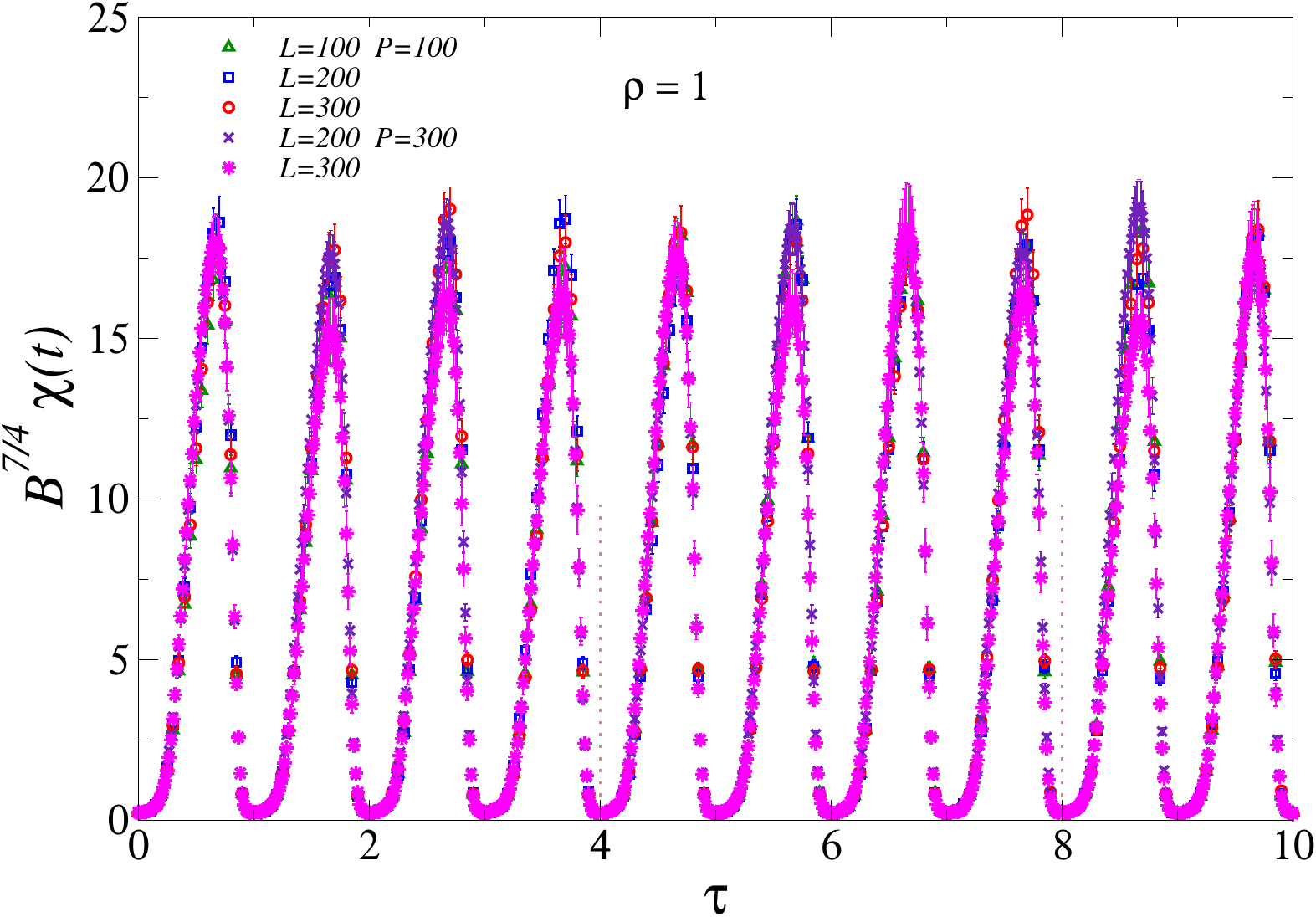}
 \includegraphics*[scale=\graphicscale]{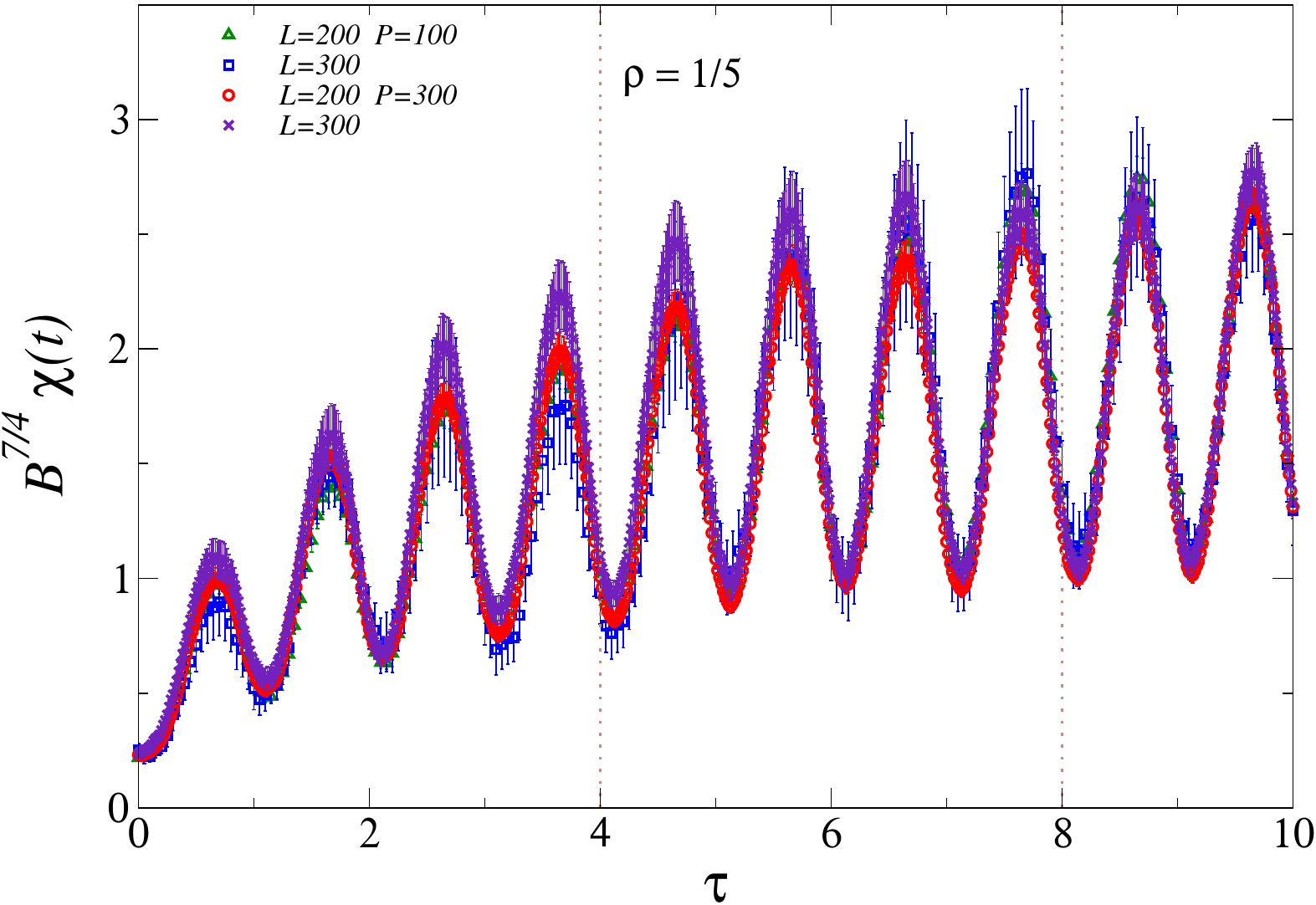}
  \caption{Rescaled magnetic susceptibility under
    periodic driving of the temperature. We report results for 
    $B^{7/4} \chi(t,B,P)$ for $\rho=1$
    (top) and $\rho=1/5$ (bottom) in the thermodynamic limit (as
    checked by the data for different sizes), and for different values
    of $P$, up to $\tau=10$.  These plots nicely confirm the dynamic
    scaling prediction, Eq.~(\ref{chitthlim}).  }
      \label{trchi}
\end{figure}
%%%%%%%%%%%%%%%%%%%%%%%%%%%%%%%%%%%%%%%%%%%%%%%%%%%%%%%%%%%%%%%%%%%%%%%%

A dynamic scaling theory can be developed as in the magnetic case. 
In finite systems of size $L$, the relevant scaling variables are 
\begin{equation}
W_e=B L^{y_t},  \qquad
W_t=t L^{-z},  \qquad
W_p=P L^{-z}. 
\end{equation}
From these expressions 
we can easily derive the dynamic scaling laws 
in the thermodynamic limit. The 
scaling variables are  $\tau=t/P$ and
\begin{eqnarray}
  \rho = W_e \, W_p^{y_t/z} = B P^{\lambda}, \qquad \lambda = y_t/z,
  \label{rhodef}
\end{eqnarray}
with $\lambda=0.4615(2)$ for the 2D Ising transition. The dynamic
scaling behavior is defined as the limits $B\to 0$, $t,P\to\infty$
keeping $\rho$ and $\tau$ fixed.  Then, since the RG dimension of the
order-parameter field is $y_\phi=(d+2-\eta)/2$, we obtain for the
time-dependent susceptibility (\ref{chitdef})
\begin{equation}
  \chi(t,B,P)\approx B^{-(d-2y_\phi)\nu} {\cal G}(\rho,\tau), 
    \label{chitthlim}
\end{equation}
where $(d-2y_\phi)\nu=(2-\eta)\nu$. At the 2D Ising transition
$(2-\eta)\nu= 7/4$. Analogously, the subtracted energy obeys the 
dynamic scaling law
\begin{equation}
  E_{se}(t,B,P)\approx B^{y_e/y_t} {\cal E}_{se}(\rho,\tau),
    \label{tenethlim}
\end{equation}
where $y_e/y_t = 2\nu-1$, thus $y_e/y_t=1$ for the 2D Ising
transition. 

To verify the dynamic scaling predictions reported above, we have
performed numerical simulations for two different values of $P$ and
two different values of $\rho$. In all cases, we have performed
simulations for different values of $L$, to identify in which range of
sizes the data can be considered as effectively in the thermodynamic
limit.

Results for the magnetic susceptibility are reported in
Fig.~\ref{trchi}.  They are in full agreement with the dynamic scaling
theory: the properly rescaled data for each given value of $\rho$---we
consider $\rho=1$ and $1/5$---and different values of $P$ fall onto a
single curve when plotted versus $\tau = t/P$, as predicted by
Eq.~(\ref{chitthlim}).  We note that some of the features already
observed in the case of the magnetic driving at $T_c$ also appear
here, for example, the full synchronization of $\chi(t)$ with the
temperature change, with a phase delay depending on $\rho$. Note also
the presence of a significant drift of the curves as $\tau$ increases,
a phenomenon we have already observed in the magnetic case for $\sigma
\lesssim 2$.

\begin{figure}[!t]
 \includegraphics*[scale=\graphicscale]{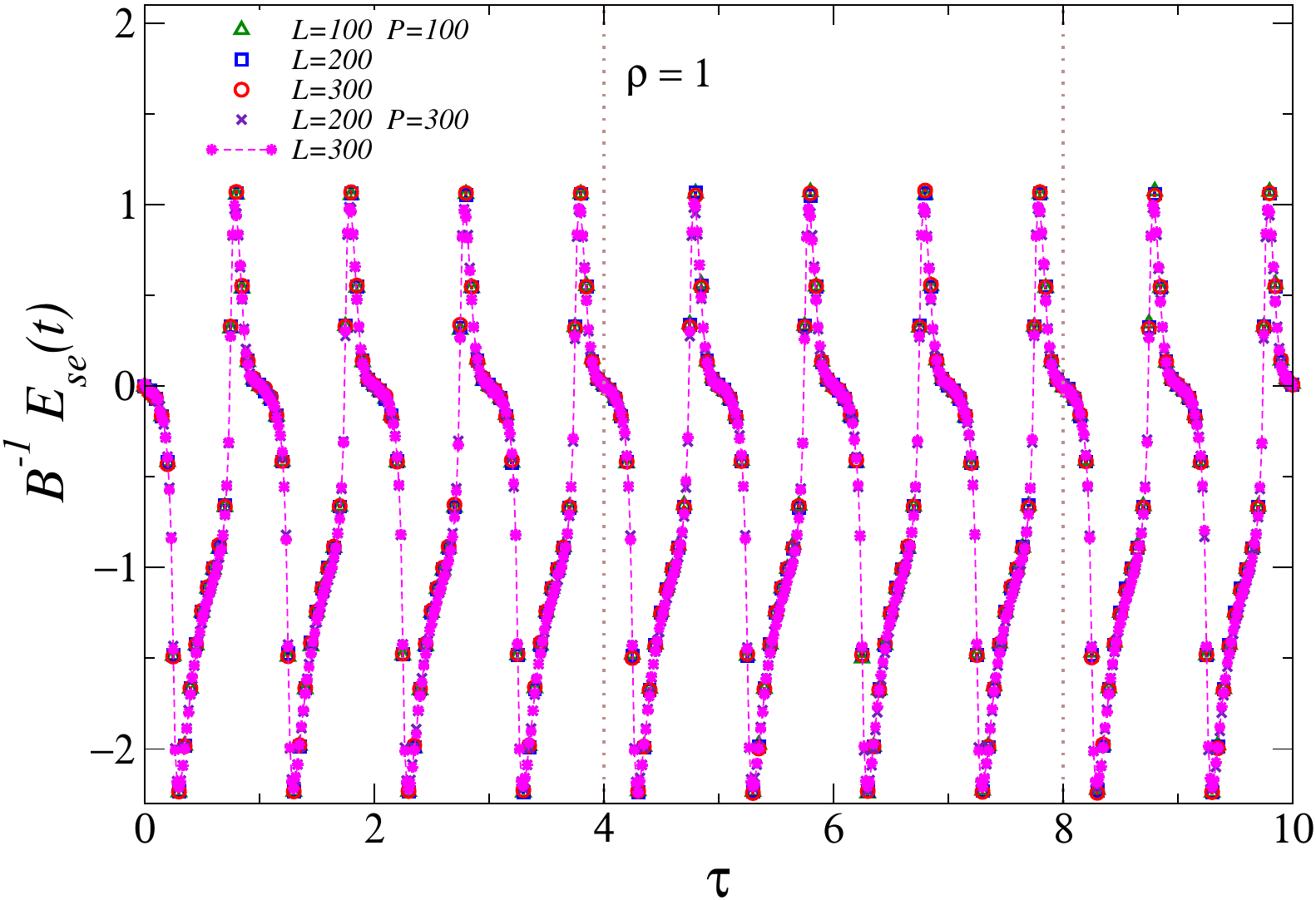}
  \includegraphics*[scale=\graphicscale]{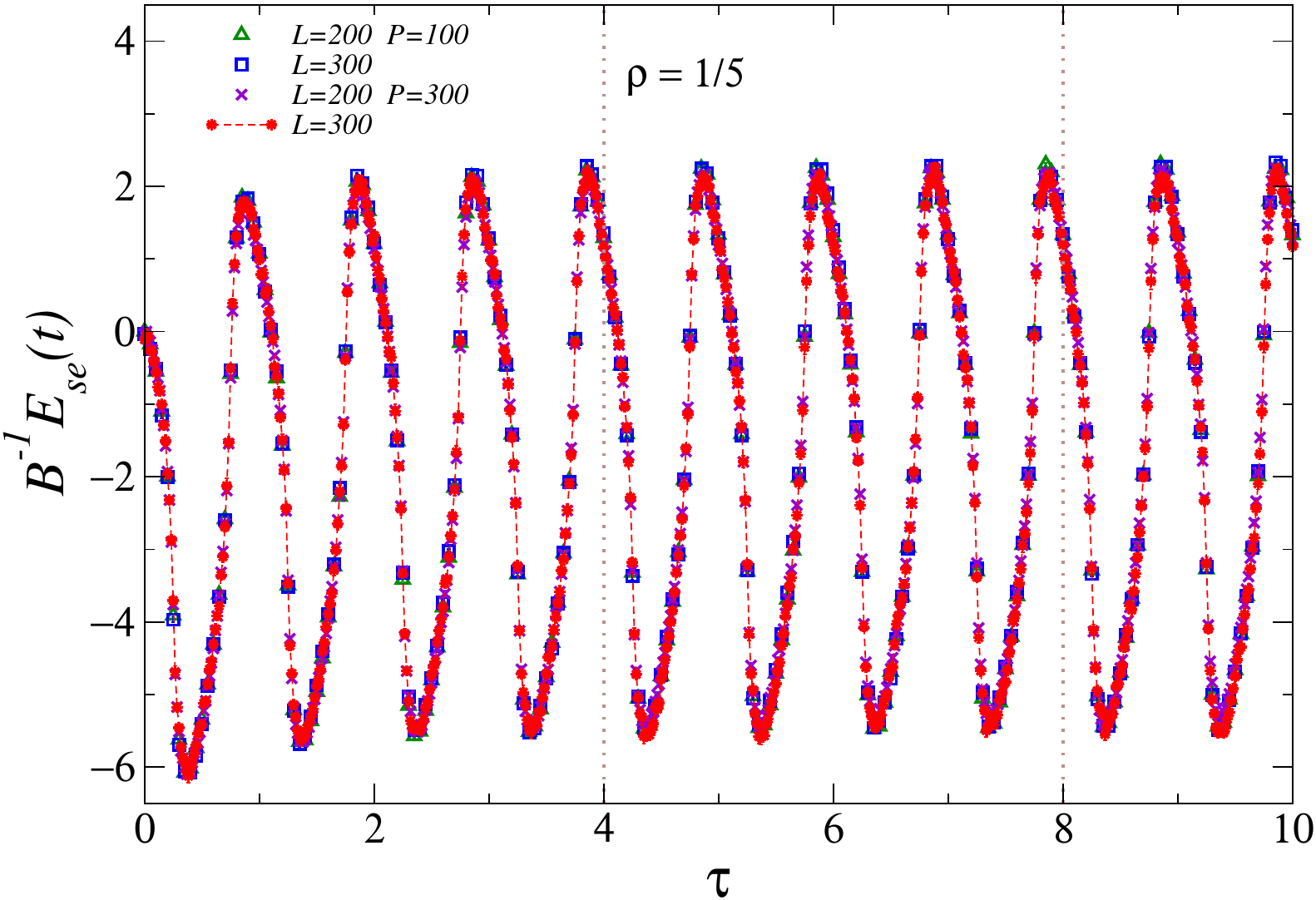}
  \caption{Rescaled subtracted bond-energy density defined in
    Eq.~(\ref{esdef2}), under the periodic driving of the temperature.
    We report results for $B^{-1} E_{se}(t,B,P)$ for $\rho=1$ (top)
    and $\rho=1/5$ (bottom) in the thermodynamic limit (as checked by
    the data for different sizes), and for different values of $P$.
    These plots confirm the dynamic scaling law,
    Eq.~(\ref{tenethlim}).}
      \label{trene}
\end{figure}

As shown in Fig.~\ref{trene}, also the data for the subtracted
bond-energy density $E_{se}(t,B,P)$, cf. Eq.~(\ref{esdef2}), show a
genuine dynamic scaling behavior.  We stress that in this case the
correct subtraction is crucial to observe dynamic scaling. Indeed, one
can easily verify that $E_s(t)$, defined in Eq.~(\ref{esdef}), does
not show dynamic scaling.

%%%%%%%%%%%%%%%%%%%%%%%%%%%%%%%%%%%%%%%%%%%%%%%%%%%%%%%%%%%%%%%%%%%%%%%%
%%%%%%%%%%%%%%%%%%%%%%%%%%%%%%%%%%%%%%%%%%%%%%%%%%%%%%%%%%%%%%%%%%%%%%%%

\section{Conclusions}
\label{conclu}

We study how periodic time-dependent perturbations of a system at a
classical (thermal) continuous phase transition affect its critical
dynamic behavior. In particular, we consider a periodic external
source coupled to the order parameter in a ferromagnetic system and
determine the resulting dynamic scaling behavior of the magnetization
and of the bond-energy density in the critical region. These results
provide a universal scaling description of the dynamic critical
behavior, extending the already ample collection of results for the
static and dynamic critical behavior of systems at continuous phase
transitions.

As a theoretical laboratory, we consider the 2D Ising model on a
square lattice at the critical temperature with a time-dependent
magnetic field $h(t)$, see Eq.~(\ref{htdef}), which drives the system
across the critical point.  The system evolves under a standard
Metropolis dynamics~\cite{Metropolis:1953am,Binder-76}, an example of
a purely relaxational dynamics (model-A dynamics~\cite{HH-77}),
starting at $t=0$ from an ensemble of negatively-magnetized
thermalized configurations at $h_i < 0$. We only consider the behavior
in the infinite-volume limit, taken at fixed Hamiltonian parameters,
which is essentially observed on lattices of size $L\gtrsim 200$ for
the set of parameters we consider.  The periodic driving leads to a
peculiar dynamic scaling behavior in terms of the scaling variables
$\sigma=A P^\kappa$ and $\tau=t/P$, cf. Eqs.~(\ref{whdef}) and
(\ref{wpdef}), where $t> 0$ is the time, and $A$ and $P$ are the
amplitude and the period of the magnetic field $h(t)$,
respectively. The (infinite-volume) magnetization and bond-energy
density have an oscillatory behavior, which is synchronized with that
of the magnetic field: They have period $P$ (or one in the variable
$\tau=t/P$), with a phase delay with respect to the magnetic field
that depends of $\sigma$ and vanishes in the limit $\sigma\to\infty$,
The center of the oscillations changes initially with time, becoming
approximately time-independent after a time interval $t_s$ that
increases with decreasing $\sigma$. The estimates of $t_s$ are
consistent with a power-law divergence, $\tau_s = t_s/P
\sim\sigma^{-u}$, with $u\approx 4$, in agreement with its expected
divergence in the limit $\sigma\to 0$.  In the large-time stationary
state the magnetization oscillates with vanishing average, recovering
the ${\mathbb Z}_2$ parity invariance broken be the starting condition
of the dynamics. Finally, we find that the amplitudes of the
oscillations of the magnetization turn out to be approximately
constant in the whole time evolution, i.e., both in the initial
transient regime and in the stationary state.

We also discuss an alternative periodic driving obtained by setting $h
= 0$ and varying periodically the temperature across the critical
point.  We determine the dynamic scaling behavior, which is again
characterized by a full synchronization of the critical modes with the
periodic driving.

We remark that the dynamic scaling theory we put forward is quite
general and thus it should also apply to systems in higher dimensions,
and, more generally, to critical statistical models driven by a
periodic external field coupled to the order parameter across a
continuous transition.  In particular, in the case of
three-dimensional Ising models subject to a periodic magnetic driving
at the critical temperature, we expect an analogous dynamic scaling,
described by Eqs.~(\ref{mtthlim}) and (\ref{enethlim}), with
corresponding scaling variables $\sigma=A P^\kappa$ with $\kappa =
y_h/z\approx 1.226$ (using the estimates of the critical exponents
reported in Refs.~\cite{PV-02,Hasenbusch-20,KPSV-16}) and $\tau=t/P$.

We finally mention that dynamic scaling should also be observed in
quantum Ising models, subject to a periodic longitudinal or transverse
field driving them across their quantum transition. Although some of
the results presented here should also apply to the quantum case,
major differences are expected. First of all, in the quantum case one
should consider the unitary quantum dynamics, with dynamic exponent
$z=1$.  Consequently, the scaling exponents and scaling functions are
expected to differ from those at classical transitions.  Moreover, in
the absence of dissipation, one expects the quantum dynamics to drive
the system outside the critical regime after a large time, because of
the continuous energy injection into the quantum system due to the
periodic driving.  A detailed analysis of the quantum case is
therefore needed, to define the correct scaling picture and to
understand under which conditions dynamic scaling can be effectively
observed (some related studies were reported in
Refs.~\cite{RSS-12,LMPPA-17}).

\end{document}